\documentclass[useAMS,usenatbib]{mn2e}
\usepackage{graphicx}
\usepackage{amsmath}
\def\be{\begin{equation}}
\def\ee{\end{equation}}
\def\ba{\begin{eqnarray}}
\def\ea{\end{eqnarray}}
\def\go{\mathrel{\raise.3ex\hbox{$>$}\mkern-14mu
             \lower0.6ex\hbox{$\sim$}}}
\def\lo{\mathrel{\raise.3ex\hbox{$<$}\mkern-14mu
             \lower0.6ex\hbox{$\sim$}}}

\def\bL{\mbox{\boldmath $L$}}
\def\bS{\mbox{\boldmath $S$}}
\def\bJ{\mbox{\boldmath $J$}}
\def\bT{\mbox{\boldmath $T$}}
\def\hatL{\hat{\mbox{\boldmath $L$}}}
\def\hatS{\hat{\!\mbox{\boldmath $S$}}}
\def\hatJ{\hat{\!\mbox{\boldmath $J$}}}

\def\rin{r_{\rm in}}
\def\rout{r_{\rm out}}
\def\Sigmain{\Sigma_{\rm in}}

\def\bOms{{\bar\Omega}_\star}
\def\brin{{\bar r}_{\rm in}}
\def\brout{{\bar r}_{\rm out}}
\def\bMstar{{\bar M}_\star}
\def\bRstar{{\bar R}_\star}
\def\bMd{{\bar M}_{\rm d}}

\def\thesd{\theta_{\rm sd}}
\def\thedb{\theta_{\rm db}}
\def\thesb{\theta_{\rm sb}}

\def\rmd{{\rm d}}
\def\rmb{{\rm b}}
\def\rms{{\rm s}}
\def\rmps{{\rm ps}}
\def\rmpd{{\rm pd}}
\def\rmsd{{\rm sd}}

\def\cN{{\cal N}}
\def\bn{{\bar n}}

\begin{document}
\title[Star-Disc-Binary Interactions]
{Star-Disc-Binary Interactions in Protoplanetary Disc Systems 
and Primordial Spin-Orbit Misalignments}
\author[D.~Lai]
{Dong Lai\thanks{Email: dong@astro.cornell.edu}\\
Center for Space Research, 
Department of Astronomy, Cornell University, Ithaca, NY 14853, USA\\}

\pagerange{\pageref{firstpage}--\pageref{lastpage}} \pubyear{2013}

\label{firstpage}
\maketitle

%%%%%%%%%%%%%%%%%%%%%%%%%%%%%%%%%%%%%%%%%%%%
\begin{abstract}
We study the interactions between a protostar and its circumstellar
disc under the influence of an external binary companion to determine
the evolution of the mutual misalignment between the stellar spin and
the disc angular momentum axes.  The gravitational torque on the disc
from an inclined binary companion makes the disc precess around the
binary axis, while the star-disc interaction torque due to the
rotation-induced stellar quadrupole tends to make the stellar spin and
the disc angular momentum axes precess around each other. A
significant star-disc misalignment angle can be generated from a small
initial value as the star-disc system evolves in time (e.g., with
decreasing disc mass) such that the two precession frequencies cross
each other. This ``secular resonance'' behaviour can be understood in
a simple, geometric way from the precession dynamics of spin and disc
angular momenta. We derive the general conditions for such resonance
crossing to occur, and find that they can be satisfied for very
reasonable protostar-disc-binary parameters. The evolution of
star-disc inclination is also significantly affected by mass accretion
from the disc onto the central star and by magnetic star-disc
interaction torques, which can either promote or reduce star-disc
misalignment, as well as by the possible damping of disc-binary
inclination due to viscous dissipation of disc warps. In general, as
long as the initial binary-disc inclination is not too small (greater
than a few degrees), a variety of star-disc misalignment angles can be
generated within the lifetimes of protoplanetary discs ($\sim
10$~Myrs). We discuss the implications of our results for the
observations of stellar spin orientations in binaries, for the
alignments/misalignments of protostellar discs and debris discs
relative to their host stars, and for the observed stellar obliquities
in exoplanetary systems. In particular, if hot Jupiters are produced
by the secular Lidov-Kozai effect induced by an external stellar
companion present in the protostellar phase, then it is likely that
``primordial'' star-disc misalignments are already generated by the
star-disc-binary interactions.  Even for systems where the Kozai
effect is suppressed, misaligned planets and hot Jupiters may still be
produced during the protoplanetray disc phase.
\end{abstract}


%%%%%%%%%%%%%%%%%%%%%%%%%%%%%%%%%%%%%%%%%%%%
\section{Introduction}

In the last few years, many exoplanetary systems containing hot
Jupiters have been be found to have high stellar obliquities, i.e.,
large misalignment angles between the spin axis of the host star and
the planetary orbital angular momentum axis (e.g., Hebrard et
al.~2008; Winn et al.~2009,2010; Triaud et al.~2010; Moutou et
al.~2011; Albrecht et al.~2012). This indicates that a large
population of hot Jupiters are formed through high-eccentricity
channels, either via dynamical planet-planet scatterings (Rasio \&
Ford 1996; Weidenschilling \& Marzari 1996; Chatterjee et al.~2008;
Juric \& Tremaine 2008) or/and secular interactions between multiple
planets, or Lidov-Kozai effect induced by a distant companion (e.g.,
Holman et al.~1997; Wu \& Murray 2003; Fabrycky \& Tremaine 2007; Wu et
al.~2007; Nagasawa et al.~2008; Naoz et al. 2011,2102; Wu \& Lithwick
2011; Katz et al.~2011).
%see Socrates et al.~2012; Dawson \& Murray-Clay 2013). 
Alternatively, Rogers et al.~(2012) suggested that the observed
stellar obliquities may result from the torques of varying directions
deposited on the radiative stellar envelope by internally generated
gravity waves.

The observed stellar obliquities may also contain contributions from 
(or be affected by) possible ``primordial'' misalignment between the 
the stellar spin and the protoplanetary disc. Several lines of evidence
suggest that such primordial misalignment is possible:

(i) In the Solar system, all major planets lie in the same plane to within
$2^\circ$, but this plane is inclined with respect to 
the solar rotational equator by $7^\circ$.

(ii) Solar-type main-sequence binaries with separations $\go
40$~AU often have rotation axis misaligned relative to the orbital
angular momentum (Hale 1994). 
%Such binaries are unlikely to have expeienced significant 
%post-formation dynamical processes.

(iii) Some binary young stellar objects (YSOs) are observed to have
circumstellar discs that are misaligned with the binary orbital plane
(e.g., Stapelfeldt et al.~1998,2003; Neuhauser et al.~2009).  
Also, several unresolved YSOs or
pre-main sequence binaries have jet axes along different directions
compared to the binary axes, suggesting misaligned discs (e.g.,
Davis, Mundt \& Eisl\"offel 1994; Roccatagliata et al.~2011).

(iv) Recently, stellar obliquities in several multiplanet systems have
been measured or constrained. While for some systems (Kepler-30,
Sanchis-Ojeda et al.~2012; KOI-94 and Kepler-25, Albrecht et al.~2013,
Hirano et al.~2012; Kepler-50 and Kepler-65, Chaplin et al.~2013), the
stellar obliquities are consistent with zero (with typical error bars
of order 10$^\circ$), highly misaligned systems have also been
found.  In particular, Huber et al.~(2013) used asteroseismology to
measure a large stellar obliquity ($40^\circ$-$55^\circ$) for
Kepler-56, a red giant star hosting two transiting coplanar planets
with orbital period of 10.5 and 21 days. A companion in a wide orbit
(as suggested by radial velocity data) may be responsible for the
observed stellar obliquity; but it is possible that the companion
was already present in the protoplanetary disc phase and produced
a primordial misalignment.  Also, there exists tentative evidence that
the Kepler-9 system, containing three transiting, roughly coplanar
planets (Holman et al.~2010), has a misaligned stellar spin (about
45$^\circ$) with respect to the planetary orbit (Walkowicz \& Basri
2013). Hirano et al.~(2014) presented several other candidate misaligned
multiplanet systems.

%has a host star that has a spin inclination (relative to 
%the line of sight) of $45^\circ\pm 10^\circ$, implying a misaligned stellar
%spin with respect to planetary orbit, although there is significant uncertainty
%in the measurement of $v\sin i$ for the star (Walkowicz \& Basri 2013).

In contrast to the above, the current observational situation with
debris disc systems seems different. In a number of debris disc
systems, the stellar spin inclination (relative to the line of sight)
$i_\star$ is similar to the debris disc inclination $i_{\rm disc}$, 
typically to within 10$^\circ$ (Watson et al.~2011; Greaves et
al.~2013; Kennedy et al.~2013). While $i_\star\approx 
i_{\rm disc}$ does not necessarily imply alignment between the spin axis and 
the disc axis, statistically these observations do suggest that 
there is no significant star-disc misalignment in these systems.

There are several proposed mechanisms for tilting a star relative to
its protoplanetary disc to produce primordial misalignment. Bate et
al.~(2010) suggested that in the chaotic star formation scenario, the
accreting gas assembled onto a protoplanetary disc may have a varying
direction of angular momentum, and thus the stellar spin direction can
be different from the disc orientation at the later phase of
accretion. Misalignment may also be produced when a protostellar
system encounters another disc/envelope system with a different
direction of rotation (Thies et al.~2011). Lai, Foucart \& Lin (2011)
showed that magnetic star-disc interaction produces a misalignment
torque between the stellar spin and the disc axis (see also Lai 1999). 
However, this torque is of the same order of magnitude as the accretion torque,
which tends to align the spin with the disc. Given the intrinsic
complexity of the magnetic accretion physics, the net effect of the
accretion and magnetic torques cannot be definitively determined at
present: All one can say is that small or modest spin-disc
misalignments may be produced in some systems; perhaps this is
responsible for the $7^\circ$ anomaly in the Solar system. To generate
larger misalignments or retrograde discs require additional external
perturbations (such as varying orientations of outer discs; see
Foucart \& Lai 2011).

Recently, Batygin (2012) suggested that the precession of a
protoplanetary disc driven by the gravitational torque from an
inclined binary companion can lead to misalignment between
the stellar spin and the disc axis. This scenario is appealing since
the same inclined binary companion is needed to make misaligned hot
Jupiters through the secular Lidov-Kozai effect. Batygin \& Adams (2013)
further considered the coupling between the rotating
central star and the precessing disc, and showed that significant
misaligment between star and disc can be produced.

Since a large fraction of protostars are in binaries, the
scenarios examined by Batygin (2012) and Batygin \& Adams (2013) are
of great interest and the mechanism of producing misalignment may be 
very robust.  On the other hand, as noted above, the 
observational situation concerning spin-disc misalignment is confusing
at present, with some protoplanetary discs showing misalignment, 
while many debris disc systems showing no significant misalignment.

In this paper we study the interactions between a protoplanetary disc,
its host star and an external binary companion on an inclined orbit in
order to understand the possibility and conditions of generating large
spin-disc misalignments. Our approach to this problem is different
from that of Batygin (2012) and Batygin \& Adams (2013), and we also
examine several additional physical effects not considered before.  In
particular, we model the star-disc-binary gravitational interactions
directly using angular momentum equations (Sections 2-3).  This allows
us to describe the rotational dynamics of the system clearly in a
physical way and see how different behaviours may arise under different
conditions (i.e., different parameters of the system, such as disc
size, rotation of the central star and the binary separation). We
elucidate the significance of ``secular resonance'' in a geometric way
(see Fig.~2 below) -- such resonance plays an important role in
producing large spin-disc misalignments.  In Section 4, we study the
effects of mass accretion and magnetic interaction between the star
and the disc, including the possibility of magnetically driven
spin-disc misalignment. We also briefly examine the effect of disc
inclination damping (relative to the binary) associated with viscous
dissipation of disc warping (Section 5). Overall, we find that all
these effects can influence the final star-disc inclination
significantly, and a variety of misalignment angles can be produced
when a protostar disc is perturbed by a binary companion. In Section 6, 
we summarize the key physical effects and results (e.g., the
conditions for resonance crossing), and discuss the implications of
our findings for stellar binaries, protoplanetary/debris 
discs and the formation of misaligned planets.

%%%%%%%%%%%%%%%%%%%%%%%%%%%%%%%%%%%%%%%%%%%%
\section{Gravitational Torques}

\subsection{Setup and Parameters of the Star-Disc-Binary System}

We consider a protostellar system consisting of a primary star
$M_\star$ surrounded by a disc and an external binary 
companion $M_\rmb$ (see Figure 1). To simplify the equations 
throughout the paper, we introduce dimensionless stellar mass, 
radius and rotation rate as
\be
{\bar M}_\star={M_\star\over 1\,M_\odot},~~
{\bar R}_\star={R_\star\over 2R_\odot},~~
\bOms={\Omega_\star\over \sqrt{GM_\star/R_\star^3}}.
\label{eq:para1}\ee
We also introduce dimensionless disc mass, inner radius and outer radius as
\be
\bMd={M_\rmd\over 0.1M_\odot},~~
\bar r_{\rm in}= {\rin\over 4R_\star},~~
\bar r_{\rm out}={\rout\over 50\,{\rm AU}}.
\label{eq:para2}\ee
Note that 
\be
\Omega_\star=\left({2\pi\over 3.3\,{\rm days}}\right)
\left({\bOms \over 0.1}\right)
\left({{\bar M}_\star\over {\bar R}_\star^3}\right)^{1/2}.
\ee
The canonical value of $\bOms$ is 0.1, 
corresponding to rotation period $P_\star=3.3$~days.
The observed $P_\star$ for pre-main-sequence stars lies in the range
between 1 and 10~days (e.g. Gallet \& Bouvier 2013). For magnetic protostars,
$\rin$ is given by the magnetosphere radius, and $\Omega_\star$ and $\rin$ are
generally related such that $\Omega_\star$ is comparable to the disc
Keplerian rotation rate at $\rin$, i.e., the spin ``fastness'' parameter,
\be
f_\star\equiv {\Omega_\star\over \Omega_{\rm k}(\rin)}=0.8
\left({\bar\Omega_\star\over 0.1}\right)\brin^{3/2},
\label{eq:fastness}\ee
is of order unity (see Section 4.1).

The disc mass decreases in time due to accretion and mass outflow,
with a lifetime of about 10~Myrs (see Section 3.3).  The shape of the
disc surface density profile also evolves in time.  For simplicity, in
this paper we assume a fixed power-law density profile (as adopted by
Batygin \& Adams 2013), 
\be 
\Sigma=\Sigmain {\rin\over r}.  
\label{eq:sigma}\ee 
Thus the total disc mass is related to $\Sigmain$ via (assuming $\rout\gg\rin$)
\be 
M_\rmd\simeq 2\pi\Sigmain\rin\rout.  
\ee 
Throughout this paper, we mostly assume
that the disc is flat (with negligible warp) and precesses like a
rigid body. The exception in Section 6, where we consider the viscous
damping of disc inclination due to disc warp. The reason the disc
behaves as a rigid body is that different regions of the disc can
communicate efficiently through bending waves and internal viscous stress
(e.g., Papaloizou \& Pringle 1983; Papaloizou \& Lin 1995; Papaloizou
\& Terquem 1995; Ogilvie 1999,2006; Bate et al.~2000;
Lubow \& Ogilvie 2000), so that only a
small disc warp is present in the disc (see Foucart \& Lai 2011,2014
and references therein). Self gravity can also enhance this
communication (Batygin 2012; Tremaine \& Davis 2013).

The disc angular momentum vector is $\bL_\rmd=L_\rmd\,\hatL_\rmd$ (where 
$\hatL_\rmd$ is the unit vector), with
\be
L_\rmd\simeq {2\over 3}M_\rmd\sqrt{GM_\star\rout}.
\ee
The stellar spin angular momentum vector is $\bS=S\,\hatS$, with 
\be
S=k_\star M_\star R_\star^2\Omega_\star.
\label{eq:spin}\ee
where $k_\star\simeq 0.2$ for fully convective stars ($n=1.5$ polytrope).
The ratio of $L_\rmd$ and $S$ is
\be
{L_\rmd\over S}=244 \left({M_\rmd\over 0.1M_\star}\right)\left({0.2\over k_\star}\right)
\left({0.1\over\bOms}\right)
\left({{\bar r}_{\rm out}\over {\bar R}_\star}\right)^{1/2}.
\label{eq:LoverS}\ee
The binary angular momentum is much larger, $L_\rmb\gg L_\rmd,\,S$.

%%%%%%%%%%%%%%%%%%%%
\begin{figure}
\begin{centering}
\vskip -0.5truecm
\includegraphics[width=12cm]{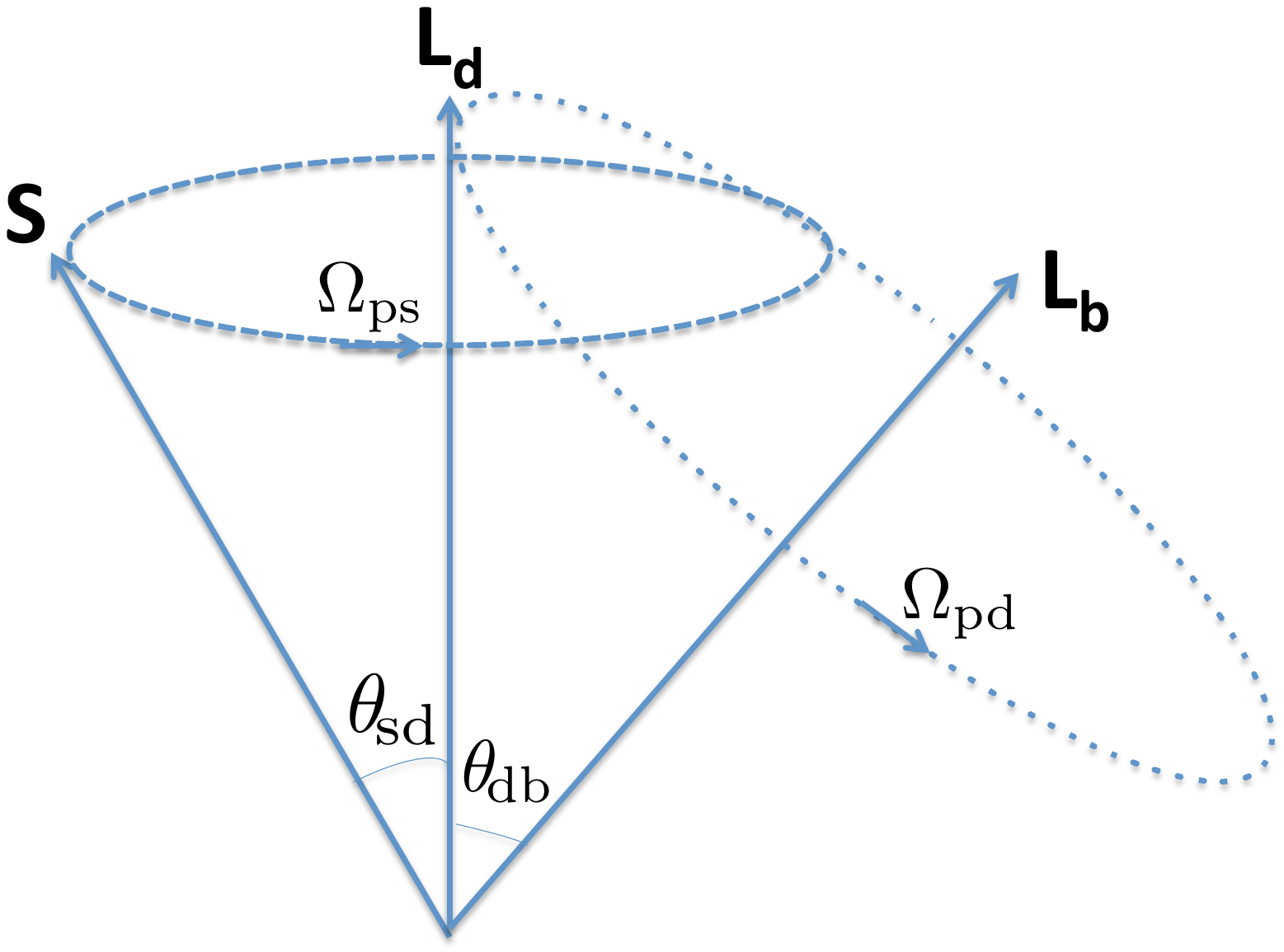}
\vskip -3.truecm
\caption{A sketch of the angular momentum axes of the star
($\bS$), circumstellar disc ($\bL_\rmd$) and external binary ($\bL_\rmb$).
The disc axis $\hatL_\rmd$ precesses around the binary axis $\hatL_\rmb$
at the rate $\Omega_\rmpd$, and the spin axis $\hatS$ precesses around
$\hatL_\rmd$ at the rate $\Omega_\rmps$ (assuming $L_\rmb\gg L_\rmd\gg S$).
}
\label{fig1}
\end{centering}
\end{figure}
%%%%%%%%%%%%%%%%%%%%%

%%%%%%%%%%%%%%%%%%%%%%%%%%%%%%%%%%%%%%%%%%
\subsection{Torques and Precession Rates}

The rotation of the star gives rise to a quadrupole moment
(the difference in moments of inertia around the two principal axes)
\be
I_3-I_1=k_q M_\star R_\star^2 \bOms^2,
\label{eq:I3I1}\ee
where $k_q\simeq 0.1$ for fully convective stars\footnote{For polytropic stellar models
(with index $n$), $k_q$ is approximately related to $k_\star$ via the relations
$k_\star=2\kappa_n/5$ and $k_q\simeq (\kappa_n^2/2)(1-n/5)$ 
[see Lai et al.~1994, eq.~(71)]. For $n=0$ (incompressible fluid), $k_n=1$
and these relations are exact. For $n=1.5$ (fully convective stars), $\kappa_n\simeq
0.51$ (see Table 1 of Lai et al.~1993), which gives $k_\star\simeq 0.2$ and 
$k_q\simeq 0.09$.}.
The gravitational torque on star from the disc is 
\be
\bT_\rms=-\int\!dM_\rmd {3G(I_3-I_1)\over 2r^3}\cos\thesd\,\hatL_\rmd\times\hatS,
\ee
where $dM_\rmd=2\pi r\Sigma dr$. Using the density profile (\ref{eq:sigma}), we find
\be
\bT_\rms =\Omega_\rmps\, 
%\hatL_\rmd\times\bS,
%\hat{\bf J}_{\rm sd}\times\bS.
\hatJ_\rmsd\times\bS.
\ee
Here $\bJ_{\rm sd}=J_{\rm sd}\,\hatJ_{\rm sd}=\bS+\bL_\rmd$ is the total 
angular momentum of the star-disc system, and $\Omega_\rmps$ is the precession rate of
the stellar spin around $\bJ_{\rm sd}$:
\be
\Omega_\rmps=-{3GM_\rmd (I_3-I_1)\over 4\rin^2\rout}
\left({J_{\rm sd}\over L_\rmd S}\right)\cos\thesd,
\ee
where $\thesd$ is the angle between $\bS$ and $\bL_\rmd$.
Using equations (\ref{eq:spin}) and (\ref{eq:I3I1}), we find
\ba
&&\Omega_\rmps
=-{3k_q\over 4k_\star}\bOms {M_\rmd\over M_\star}{\sqrt{GM_\star 
R_\star^3}\over \rin^2\rout}
\left({J_{\rm sd}\over L_\rmd}\right)\cos\thesd 
\nonumber\\
&&\qquad =-4.86\times 10^{-5}\left({2k_q\over k_\star}\right)
\left({M_\rmd\over 0.1M_\star}\right)
\left({\bOms\over 0.1}\right){1\over \brin^{2}\brout}\nonumber\\
&&\qquad\quad \times \left({\bMstar\over\bRstar}\right)^{\!\! 1/2}
\left({J_{\rm sd}\over L_\rmd}\right)\cos\thesd 
\left({2\pi\over {\rm yr}}\right).
\label{eq:Omegaps}
\ea
Note that for $L_\rmd\gg S$, we have $J_{\rm sd}\simeq L_\rmd$ and 
$\Omega_\rmps$ is simply the precession rate of $\bS$ around $\bL_\rmd$
(see Figure 1).

The torque on the disc from the external binary companion (mass
$M_\rmb$ and separation $a_\rmb$) is
\ba
&&\bT_\rmd=-\int\!dM_\rmd {3GM_\rmb r^2\over 4 a_\rmb^3}
\cos\thedb\,\hatL_\rmb\times\hatL_\rmd\nonumber\\
&&\qquad = -{GM_\rmb M_\rmd \rout^2\over 4 a_\rmb^3}\cos\thedb\,\hatL_\rmb\times
\hatL_\rmd \nonumber\\
&&\qquad \equiv \Omega_\rmpd \hatL_\rmb\times \bL_\rmd,
\ea
where $\hatL_\rmb$ is the unit vector along the binary angular momentum axis,
and $\thedb$ is the angle between $\hatL_\rmd$ and $\hatL_\rmb$.
The precession rate of the disc axis $\bL_\rmd$ around the binary 
axis $\bL_\rmb$ is then
\ba
&& \Omega_\rmpd=-{3M_\rmb\over 8M_\star}\left({GM_\star\rout^3\over a_\rmb^6}
\right)^{\!\! 1/2}\!\!\cos\thedb \nonumber\\
&&\qquad = -4.91\times 10^{-6}\left(\!{M_\rmb\over M_\star}\!\right)
\bMstar^{1/2}\brout^{3/2}\left({a_\rmb\over 
300\,{\rm AU}}\right)^{\!\!-3}\nonumber\\
&&\qquad\quad \times \cos\thedb\left({2\pi\over {\rm yr}}\right).
\label{eq:Omegapd}
\ea

Finally, the binary companion also exerts a torque on the oblate star,
given by
\be
\bT_\rms'=-{3GM_\rmb(I_3-I_1)\over 2a_\rmb^3}(\hatL_\rmb\cdot\hatS)\left(\hatL_\rmb\times\hatS
\right).
\ee
But $|\bT_\rms'|$ is many orders of magnitude smaller than
$|\bT_\rms|$, and will be neglected in this paper.

The ratio between $\Omega_\rmps$ and $\Omega_\rmpd$ is of great importance
for the spin evolution (Section 3) and is given by
\ba
&&{\Omega_\rmps\over\Omega_\rmpd}=9.9\left({M_\rmb\over M_\star}\right)^{\!\!\!-1}
\!\!\left(\!{M_\rmd\over 0.1M_\star}\!\right)
{1\over \brout^{5/2}\brin^2}\!\left(\!{\bOms\over 0.1}\!\right)
\!\left(\!{a_\rmb\over 300\,{\rm AU}}\!\right)^3\nonumber\\
&&\qquad\quad\times {1\over \bRstar^{1/2}}
\!\left(\!{2k_q\over k_\star}\!\right)
\!\left(\!{J_{\rm sd}\over L_\rmd}\!\right)
\!\left(\!{\cos\thesd\over\cos\thedb}\!\right).
\label{eq:ratio}\ea

%%%%%%%%%%%%%%%%%%%%%%%%%%%%%%%%%%%%%%%%%%%%
\section{Evolution of Spin Direction Due to Gravitational Torques}

The gravitational torques discussed in Section 2 do not change the magnitudes of
$\bL_\rmd$ and $\bS$. So the evolution equations of the spin and disc axes are
\ba
&&{d\,\hatS\over dt}=\Omega_\rmps\,\hatJ_\rmsd\times\hatS,\label{eq:dSdt}\\
&&{d \hatL_\rmd\over dt}=\Omega_\rmpd\,\hatL_\rmb\times\hatL_\rmd
+\Omega_\rmps\,\hatJ_\rmsd\times\hatL_\rmd.\label{eq:dLdt}
\ea
Since $L_\rmb\gg L_\rmd$ and $L_\rmb\gg S$, we assume that $\hatL_\rmb$ is a constant
vector throughout this paper. Also, for $L_\rmd\gg S$ [see eq.~(\ref{eq:LoverS})],
$\bJ_\rmsd\simeq \bL_\rmd$ and equations (\ref{eq:dSdt})-(\ref{eq:dLdt}) reduce to
\ba
&&{d\,\hatS\over dt}\simeq \Omega_\rmps\,\hatL_\rmd\times\hatS,\label{eq:dSdt2}\\
&&{d \hatL_\rmd\over dt}\simeq \Omega_\rmpd\,\hatL_\rmb\times\hatL_\rmd \qquad\quad
({\rm for}~L_\rmd\gg S).\label{eq:dLdt2}
\ea
Our analysis below will be based on equations (\ref{eq:dSdt2})-(\ref{eq:dLdt2}).

%%%%%%%%%%%%%%%%
\subsection{Limiting Cases: $|\Omega_\rmps|\gg |\Omega_\rmpd|$
and $|\Omega_\rmps|\ll |\Omega_\rmpd|$}

While equations (\ref{eq:dSdt2})-(\ref{eq:dLdt2}) can be integrated numerically
in general, the solutions in two limiting cases are intuitively expected
(see Section 3.2):

(i) For $|\Omega_\rmps|\gg |\Omega_\rmpd|$, the vector $\hatS$ rapidly processes
around the slowly changing $\hatL_\rmd$. Thus we expect $\hatS$ to adiabatically follow
$\hatL_\rmd$ with $\thesd\approx$ constant. The fractional variation of
$\thesd$ is of order $|\Omega_\rmpd|/|\Omega_\rmps|$.

(ii) For $|\Omega_\rmps|\ll |\Omega_\rmpd|$, the vector $\hatS$ cannot
keep up with the rapidly changing $\hatL_\rmd$. In effect, $\hatS$ will process
around $\hatL_\rmb$ with the rate $\Omega_\rmps\cos\thedb$. That is, after averaging 
over time $P_\rmpd=2\pi/|\Omega_\rmpd|$, the spin axis satisfies the equation
\be
{d\langle\,\hatS\rangle\over dt}\simeq \Omega_\rmps\cos\thedb\,\hatL_\rmb\times
\langle\,\hatS\rangle.
\ee
Thus, in this limit, we expect that the angle between $\hatS$ and $\hatL_\rmb$,
$\thesb\approx$ constant, with the fractional variation
of order $|\Omega_\rmps|/|\Omega_\rmpd|$.

%%%%%%%%%%%%%%%%
\subsection{Evolution Equations for $\thesd$ and $\thesb$}

The statements in the last subsection can be justified from the evolution
equations for $\thesd$ and $\thesb$, which we now derive from equations
(\ref{eq:dSdt2})-(\ref{eq:dLdt2}). 

In the frame rotating around $\hatL_\rmb$ at the rate $\Omega_\rmpd$,
the vector $\hatL_\rmd$ is constant, and the spin vector evolves according to
\be
\left({d\,\hatS\over dt}\right)_{\rm rot}\simeq 
\left(\Omega_\rmps\,\hatL_\rmd-\Omega_\rmpd\hatL_\rmb\right)
\times\hatS.\label{eq:dSdt3}
\ee
Thus, in this rotating frame, $\hatS$ processes around the axis $\hatL_e$ with the
rate $\Omega_e$, where
\ba
&&\Omega_{\rm e}\hatL_{\rm e}=\Omega_\rmps\hatL_\rmd-\Omega_\rmpd\hatL_\rmb,\label{eq:Le}\\
&&\Omega_{\rm e}=-\left(\Omega_\rmps^2+\Omega_\rmpd^2-2\Omega_\rmps\Omega_\rmpd\cos\thedb
\right)^{1/2}.\label{eq:Omegae}
\ea

%%%%%%%%%%%%%%%%%%%%
\begin{figure}
\begin{centering}
\vskip -0.5truecm
\includegraphics[width=12cm]{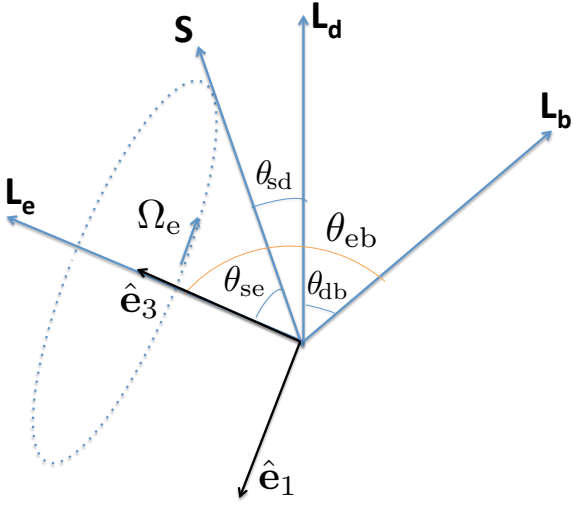}
\vskip -1.5truecm
\caption{A sketch of the angular momentum axes of the star
($\bS$), circumstellar disc ($\bL_\rmd$), and external binary ($\bL_\rmb$).
In the frame rotating around $\bL_\rmb$ with rate $\Omega_\rmpd$ (see Fig.~1), 
the vector $\bL_\rmd$ is constant (assuming $L_\rmd \gg S$),
and the stellar spin $\bS$ precesses around the vector 
$\hatL_{\rm e}$ [defined by eq.~(\ref{eq:Le})] at the rate $\Omega_{\rm e}$.
For $|\Omega_\rmps|\gg |\Omega_\rmpd|$, we have 
$\hatL_{\rm e}\simeq \hatL_\rmd$, thus $\thesd\simeq$~constant.
For $|\Omega_\rmps|\ll |\Omega_\rmpd|$, we have 
$\hatL_{\rm e}\simeq \hatL_\rmb$, thus
$\thesb\simeq$~constant. For $\Omega_\rmps=\Omega_\rmpd$ 
(``secular resonance''), $\hatL_{\rm e}\propto \hatL_\rmd-\hatL_\rmb$ deviates
significantly from $\hatL_\rmd$ even for small initial $\thesd$ and $\thedb$;
thus a large $\thesd$ can be produced as $\hatS$ precesses around $\hatL_{\rm e}$.
The Cartesian basis $(\hat{\bf e}_1,\hat{\bf e}_2,\hat{\bf e}_3)$ is used in 
eqs.~(\ref{eq:sx})-(\ref{eq:sz}). 
}
\label{fig2}
\end{centering}
\end{figure}
%%%%%%%%%%%%%%%%%%%%%

It is now easy to see the results stated in Section 3.1
(see Fig.~2):
(i) For $|\Omega_\rmps|\gg |\Omega_\rmpd|$, we have 
$\hatL_{\rm e}\simeq \hatL_\rmd$, and $\hatS$ precesses around
$\hatL_\rmd$ with constant $\thesd$.
(ii) For $|\Omega_\rmps|\ll |\Omega_\rmpd|$, we have 
$\hatL_{\rm e}\simeq \hatL_\rmb$ (with a possible sign difference), 
and $\hatS$ precesses around $\hatL_\rmb$ with constant $\thesb$.

If $\Omega_\rmps$ and $\Omega_\rmpd$ do not vary in time,
or more generally, if the rate of change of
$|\Omega_e|\hatL_e$ is much less than $|\Omega_e|$, then equation (\ref{eq:dSdt3})
can be easily solved. Suppose at $t=0$, the vector $\hatS$ lies in the same plane as
$\hatL_\rmd$ and $\hatL_\rmb$ (see Fig.~2). 
Define the basis vectors $\hat{\bf e}_3=\hatL_{\rm e}$,
$\hat{\bf e}_2\propto \hatL_\rmb\times\hatL_\rmd$, and 
$\hat{\bf e}_1=\hat{\bf e}_2\times \hat{\bf e}_3$. Then at
$t>0$, the components of $\hatS$ along these bases are
\ba
&&{\hat S}_1\simeq -\sin\theta_{\rm se}\cos\Phi_{\rm se}(t), \label{eq:sx}\\
&&{\hat S}_2\simeq -\sin\theta_{\rm se}\sin\Phi_{\rm se}(t), \label{eq:sy}\\
&&{\hat S}_3\simeq \cos\theta_{\rm se}, \label{eq:sz}
\ea
where 
\be
\Phi_{\rm se}(t)=\int_0^t\!dt\,\Omega_{\rm e},
\ee
and $\theta_{\rm se}$ is the angle between $\hatS$ and
$\hatL_{\rm e}$, and is determined by its value at $t=0$, i.e., 
$\theta_{\rm se}=\theta_{\rm se,0}=
\theta_{\rm eb}-\theta_{\rm sd,0}-\thedb$,
with $\theta_{\rm eb}$ the angle bewteen $\hatL_e$ and $\hatL_\rmb$.

From equations (\ref{eq:dSdt2})-(\ref{eq:dLdt2}), we have
\ba
{d\over dt}\cos\thesd &\simeq &\hatS\cdot\left(\Omega_\rmpd\,\hatL_\rmb\times\hatL_\rmd\right)
\nonumber \\
&\simeq &-\Omega_\rmpd\sin\thedb\sin\theta_{\rm se}\sin\Phi_{\rm se}(t),
\label{eq:dthesd}\ea
where in the second equality we have used equation (\ref{eq:sy}).
Similarly, from equation (\ref{eq:dSdt2}), we have
\ba
{d\over dt}\cos\thesb &\simeq&\hatL_\rmb\cdot\left(\Omega_\rmps\,\hatL_\rmd\times\hatS\right)
=\Omega_\rmps\,\hatS\cdot\left(\hatL_\rmb\times\hatL_\rmd\right)\nonumber \\
&\simeq& -\Omega_\rmps\sin\thedb\sin\theta_{\rm se}\sin\Phi_{\rm se}(t).
\label{eq:dthesb}\ea

For constant $\Omega_\rmpd$ and $\Omega_\rmps$, equations (\ref{eq:dthesd})
-(\ref{eq:dthesb}) give
\ba
&&\!\!\!\cos\thesd-\cos\theta_{\rm sd,0}\simeq {\Omega_\rmpd\over\Omega_e}
\sin\thedb\sin\theta_{\rm se}\left(\cos\Omega_e t-1\right),\\
&&\!\!\!\cos\thesb-\cos\theta_{\rm sb,0}\simeq {\Omega_\rmps\over\Omega_e}
\sin\thedb\sin\theta_{\rm se}\left(\cos\Omega_e t-1\right).
\ea
With these two equations, the limiting cases of Section 3.1 can be understood
precisely:

(i) For $|\Omega_\rmps|\gg |\Omega_\rmpd|$, we have 
$\hatL_{\rm e}\simeq \hatL_\rmd$, $|\Omega_e|\simeq |\Omega_\rmps|$ and 
$|\theta_{\rm se}|\simeq |\thesd|$, so 
$\thesd\approx$ constant, with variation $\Delta\thesd\sim 
(\Omega_\rmpd/\Omega_\rmps) \sin\thedb$.

(ii) For $|\Omega_\rmps|\ll |\Omega_\rmpd|$, 
$\hatL_{\rm e}\simeq \hatL_\rmb$, $|\Omega_e|\simeq |\Omega_\rmpd|$ and 
$|\theta_{\rm se}|\simeq |\thesb|$, so 
$\thesb\approx$ constant, with variation $\Delta\thesb\sim 
(\Omega_\rmps/\Omega_\rmpd)\sin\thedb$.

%%%%%%%%%%%%%%%%
\subsection{General Cases: Evolving Discs}

We now consider the cases when the parameters of disc-star system
evolve in time. In general, the stellar radius and spin can change, as
well as the disc inner and outer radii, all on timescales of order a
few Myrs. Here, for concreteness, we only consider the evolution of the
disc mass $M_\rmd$, which directly affects the spin precession rate
$\Omega_\rmps$.  In this section we neglect mass accretion and other
effects on the stellar spin (see Section 4), and we also neglect any
direct effect of changing disc mass on the disc orientation.  We
assume that the disc mass evolves in time according to (Batygin \&
Adams 2013)
\be
M_\rmd={M_{\rm d0}\over 1+t/\tau}.
\ee
For our canonical parameters, we choose $M_{\rm d0}=0.1M_\odot$ and $\tau=0.5$~Myrs.

%%%%%%%%%%%%%%%%%%%%
\begin{figure}
\begin{centering}
\vskip -0.7truecm
\includegraphics[width=8.3cm]{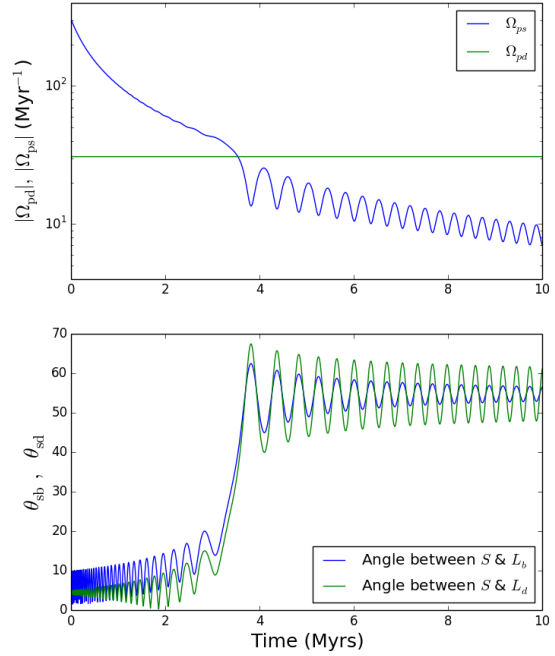}
\vskip -0.7truecm
\caption{Evolution of the stellar spin direction due to interactions
  with circumstellar disc and external binary companion. The upper
  panel shows the disc precession rate $|\Omega_\rmpd|$ and the
  stellar precession rate $|\Omega_\rmps|$, while the lower panel
  shows the spin-disc angle $\thesd$ and spin-binary angle $\thesb$. The
  initial values for various angles are: disc-binary $\thedb=5^\circ$
  (which is constant in time when $L_\rmd\gg S$), spin-disc
  $\thesd=5^\circ$, spin-binary $\thesb=10^\circ$. The initial disc
  mass is $M_{\rm d0}=0.1M_\star$. The other parameters are fixed at
  their canonical values shown in equations (\ref{eq:Omegaps}) and
  (\ref{eq:Omegapd}): $2k_q/k_\star=1$, $\bar\Omega_\star=0.1$,
   $\brin=\brout=1$, ${\bar M}_\star=\bar R_\star=1$, $M_\rmb=M_\star$, and
   $a_\rmb=300$~AU.}
\label{fig3}
\end{centering}
\end{figure}
%%%%%%%%%%%%%%%%%%%%%

%%%%%%%%%%%%%%%%%%%%
\begin{figure}
\begin{centering}
\vskip -0.7truecm
\includegraphics[width=8.3cm]{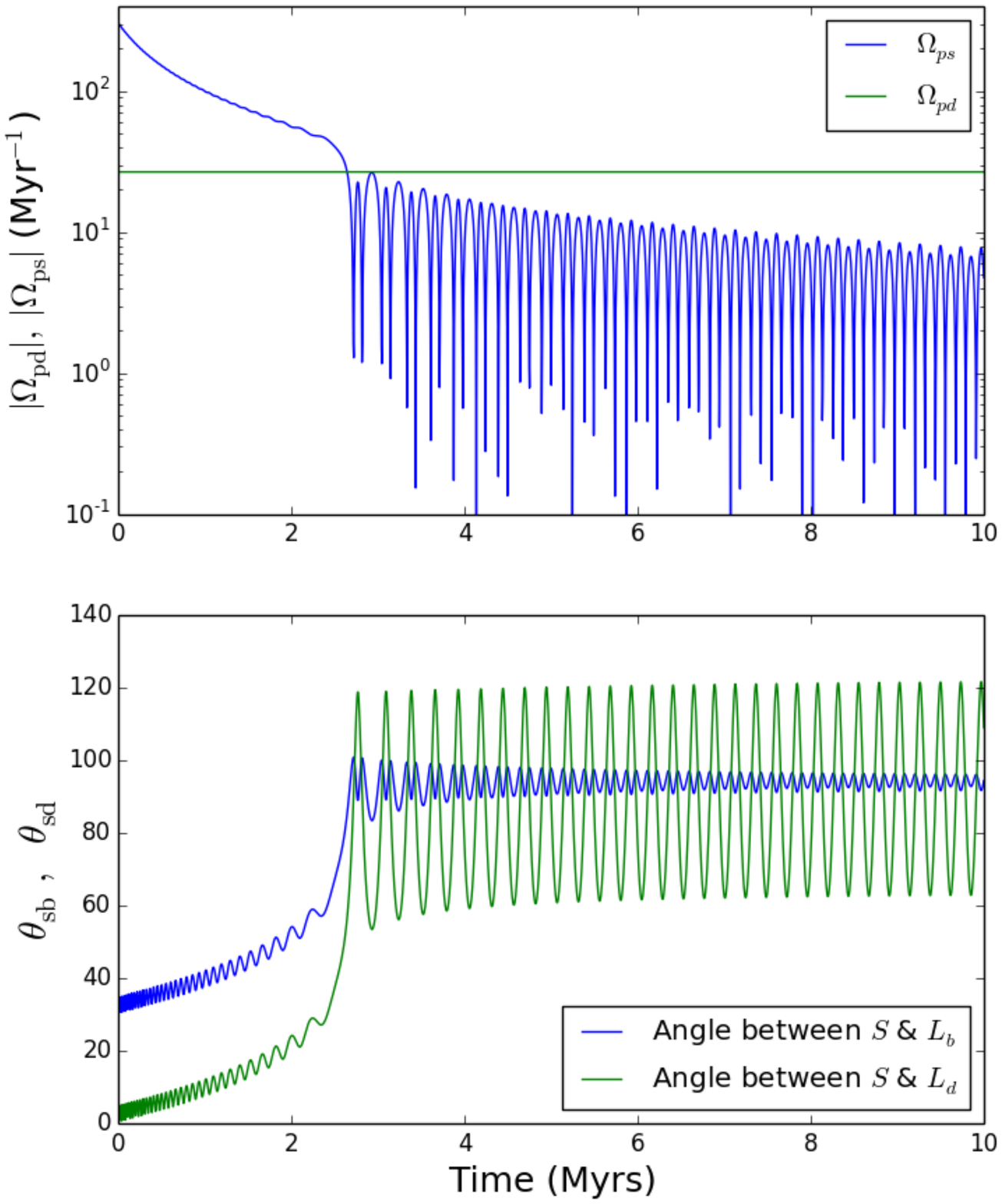}
\vskip -0.7truecm
\caption{Same as Fig.~3, except
for the initial disc-binary angle $\thedb=30^\circ$ and 
spin-binary angle $\thesb=35^\circ$.}
\label{fig4}
\end{centering}
\end{figure}
%%%%%%%%%%%%%%%%%%%%%

%%%%%%%%%%%%%%%%%%%%
\begin{figure}
\begin{centering}
\vskip -0.7truecm
\includegraphics[width=8.3cm]{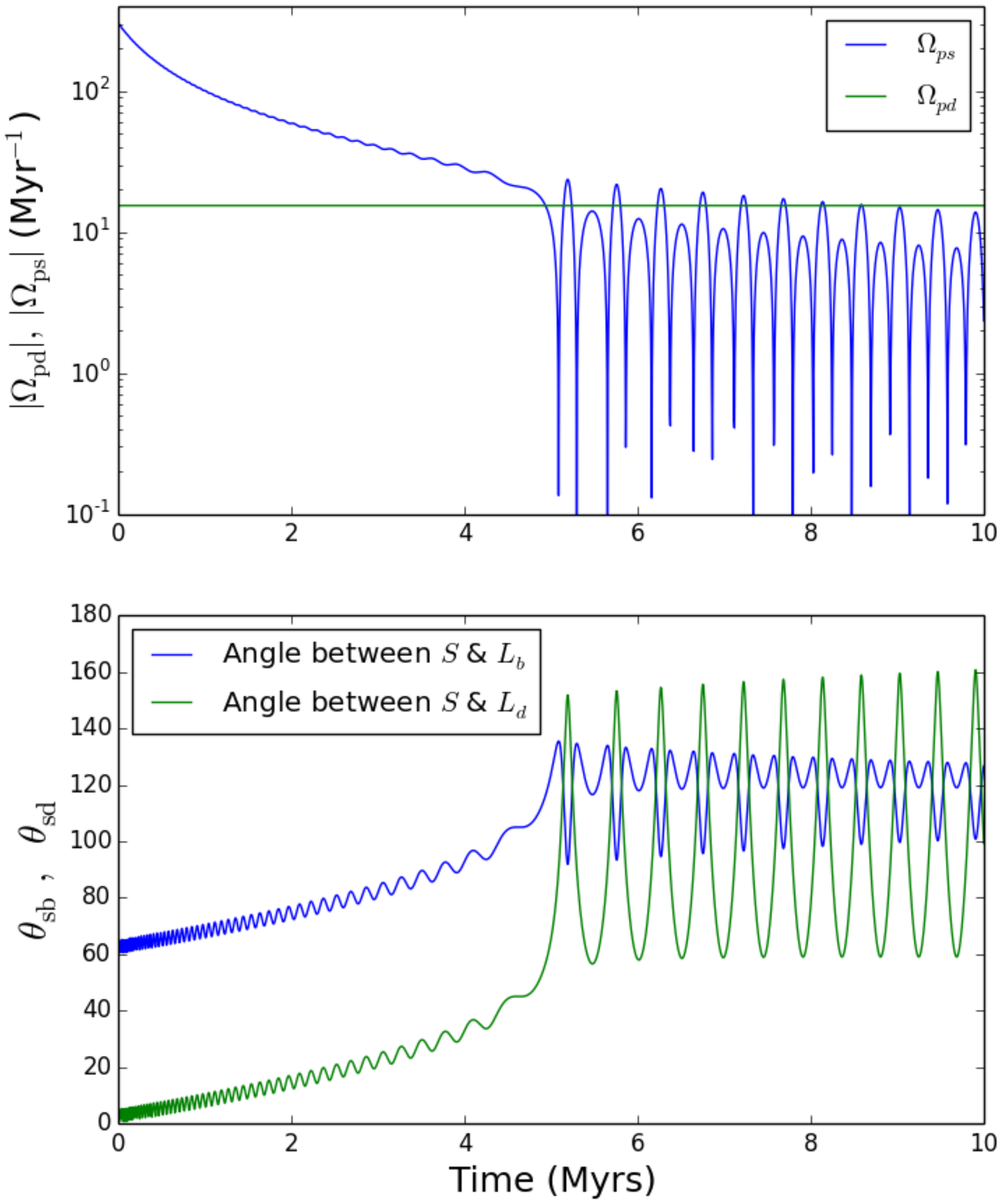}
\vskip -0.7truecm
\caption{Same as Fig.~3, except for
the initial disc-binary angle $\thedb=60^\circ$ and 
spin-binary angle $\thesb=65^\circ$.}
\label{fig5}
\end{centering}
\end{figure}
%%%%%%%%%%%%%%%%%%%%%

%%%%%%%%%%%%%%%%%%%%
\begin{figure}
\begin{centering}
\vskip -0.7truecm
\includegraphics[width=8.3cm]{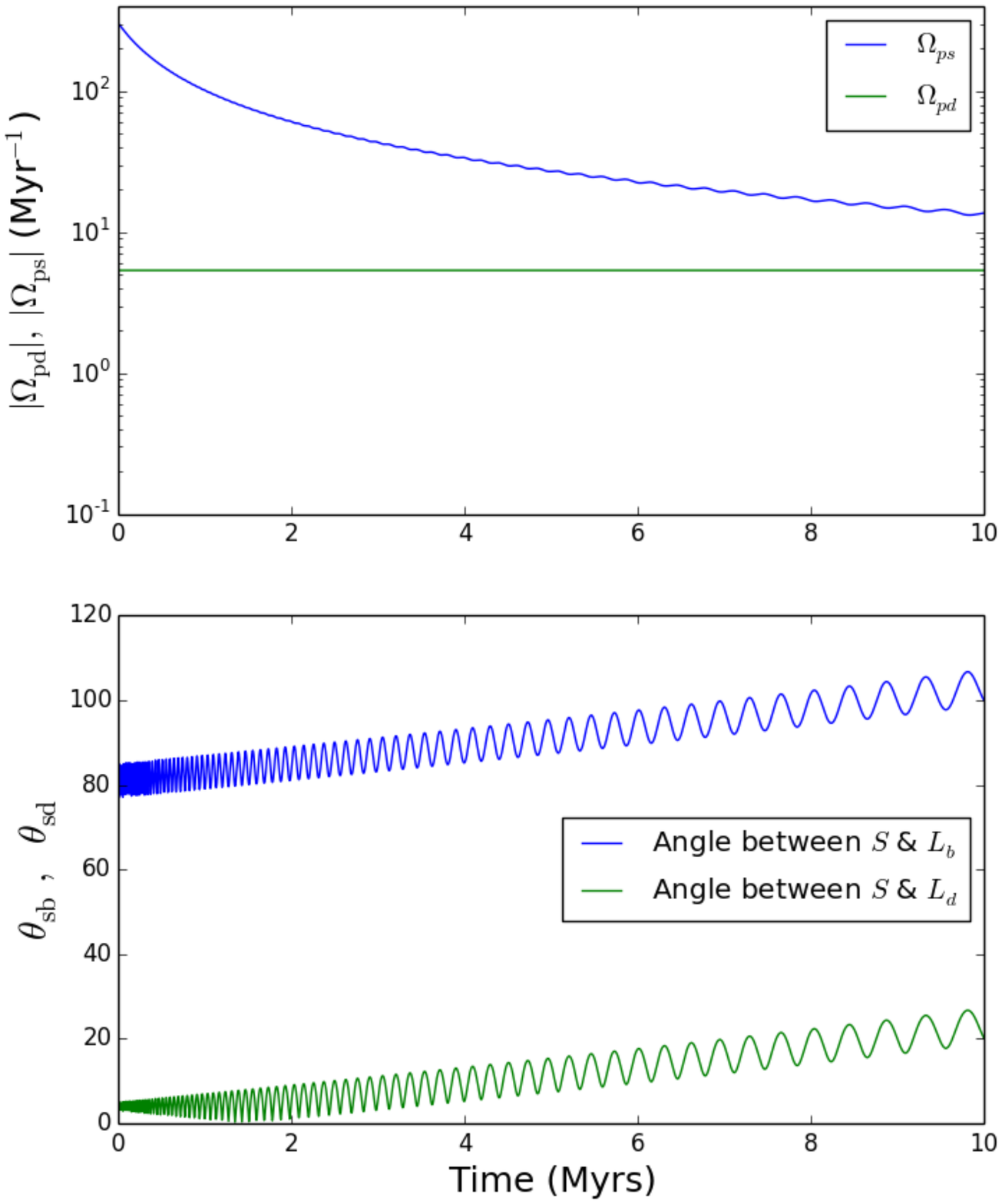}
\vskip -0.7truecm
\caption{Same as Fig.~3, except for
the initial disc-binary angle $\thedb=80^\circ$ and 
spin-binary angle $\thesb=85^\circ$.}
\label{fig6}
\end{centering}
\end{figure}
%%%%%%%%%%%%%%%%%%%%%

%%%%%%%%%%%%%%%%%%%%
\begin{figure}
\begin{centering}
\vskip -0.5truecm
\hskip -4.5truecm
\includegraphics[width=13cm]{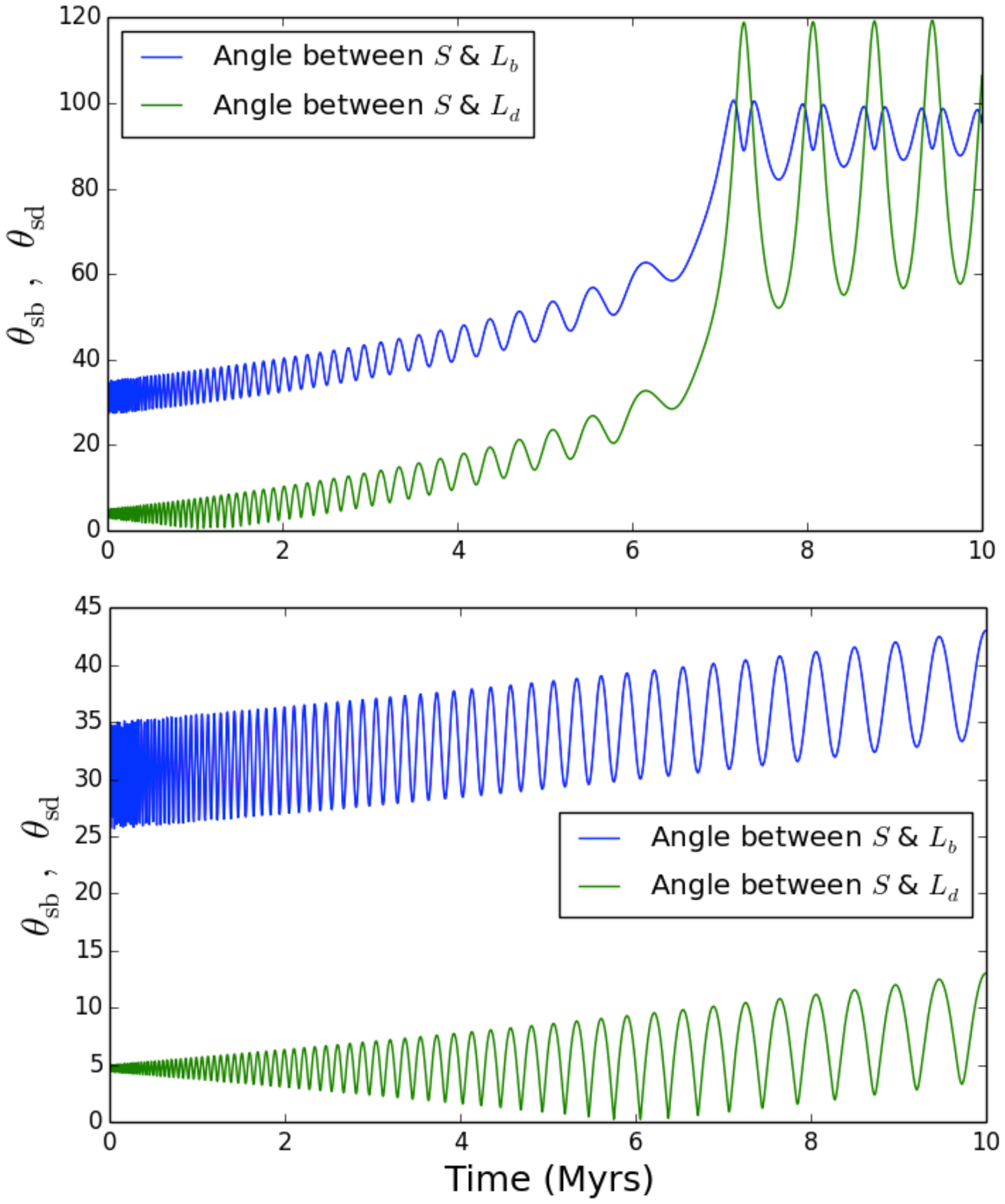}
\vskip -0.5truecm
\caption{Evolution of the stellar spin direction due to interactions with circumstellar
disc and external binary companion. The parameters are the same as in
Fig.~4 (i.e., the initial $\thedb=30^\circ$ and $\thesb=35^\circ$), except 
$a_\rmb=400$~AU (upper panel) and $a_\rmb=600$~AU (lower panel).}
\label{fig7}
\end{centering}
\end{figure}
%%%%%%%%%%%%%%%%%%%%%

Figures \ref{fig3}-\ref{fig6} show the evolution of the stellar spin direction
relative to the disc axis and to the binary axis for different initial
disc-binary angle $\thedb$ ($=5^\circ,30^\circ,60^\circ,80^\circ$). Note that this
angle does not change in time when $L_\rmb\gg L_\rmd\gg S$. The initial spin-disc
angle $\thesd$ is fixed at $5^\circ$, and $\hatS,\hatL_\rmd,\hatL_\rmb$ lie in the same
plane at $t=0$.  The other parameters assume their canonical values
made explicit in equations (\ref{eq:Omegaps}) and (\ref{eq:Omegapd}): 
$M_\star=1\,M_\odot$, $R_\star=2R_\odot$, $\bar\Omega_\star=0.1$, $k_\star=2k_q$,
$\rin=4R_\star$, $\rout=50$~AU, $M_\rmb=M_\star$ and $a_\rmb=300$~AU.
We see that at early time, when $M_\rmd$ is large, the inequality $|\Omega_\rmps|
\gg |\Omega_\rmpd|$ is satisfied, and $\hatS$ adiabatically tracks $\hatL_\rmd$,
with $\thesd$ approximately holding constant. Unless $|\cos\thedb|$ is too small 
(see Fig.~\ref{fig6} for the case of $\thedb=80^\circ$), 
which implies very small $|\Omega_\rmpd|$, the system will evolve
to the regime $|\Omega_\rmpd|\go |\Omega_\rmps|$ within 10~Myrs, 
and large $\thesd$ can be developed. As expected (see Section 3.1), when
$|\Omega_\rmpd|\gg |\Omega_\rmps|$, the stellar spin effectively precesses around
the binary axis $\hatL_\rmb$, with $\thesb$ approximately constant.
The detailed evolution of the spin direction is complicated by the fact that 
$\Omega_\rmps$ crosses zero at $\thesd=90^\circ$.

The evolution of spin direction depends sensitively on the binary
separation $a_\rmb$.  Figure~\ref{fig7} depicts some examples for
$a_\rmb=400$ and 600~AU (the corresponding case of $a_\rmb=300$~AU is
shown in Fig.~4).  Obviously, for distant companion ($a_\rmb\go
10^3$~AU), the condition $|\Omega_\rmps|\gg |\Omega_\rmpd|$ is always
satisfied, and the stellar spin can adiabatically follow the
slowly-precessing disc and the angle $\thesd$ retains its initial
value.

\subsection{Understanding ``Secular Resonance''}

The fact that significant change in $\thesd$ occurs when
$|\Omega_\rmps|\sim |\Omega_\rmpd|$ is a signature of ``secular
resonance'' (Batygin \& Adams 2013).  The significance of this
resonance can be understood geometrically from Fig.~2: For
$\Omega_\rmps=\Omega_\rmpd$, the vector $\hatL_{\rm e}\propto
\hatL_\rmd-\hatL_\rmb$ deviates significantly from $\hatL_\rmd$ even
for small initial $\thesd$ and $\thedb$; thus $\thesd$ oscillates from
its initial (possibly small) value to a very large value as $\hatS$ precesses around
$\hatL_{\rm e}$.  In the cases where the ratio
$\Omega_\rmps/\Omega_\rmpd$ varies in time and transitions from above
unity to below unity (as in the examples depicted in Section 3.3), this
oscillation of $\thesd$ is incomplete, but a large value
of $\thesd$ can be generated as the system crosses the resonance.

The actual change of $\thesd$ associated with resonance crossing depends
on how fast the system evolves through the resonance, or in the examples
of Section 3.3, how fast the disc mass decreases. Consider two limiting cases:

(i) If $|\Omega_\rmps/\Omega_\rmpd|$ evolves from $\gg 1$ to $\ll 1$ very slowly,
at an rate smaller than $|\Omega_e|$ for all times, i.e., 
if $|\dot\Omega_e/\Omega_e|\ll |\Omega_e|$, then $\hatS$ will adiabatically precess
around $\hatL_{\rm e}$ with a constant $\theta_{\rm se}$. Thus, if the initial
spin-disc angle is $\theta_{\rm sd,0}$, we have 
$\theta_{\rm se}=\theta_{\rm sd,0}$
(since $\hatL_{\rm e}=\hatL_\rmd$ initially); in the end 
(when $|\Omega_\rmps/\Omega_\rmpd|\ll 1$), we have 
$\hatL_{\rm e}=\hatL_\rmb$ and $\thesb=\theta_{\rm se}=\theta_{\rm sd,0}$,
with $\thesd$ oscillating between $|\theta_{\rm sd,0}-\thedb|$
and $\theta_{\rm sd,0}+\thedb$.

(ii) If $|\Omega_\rmps/\Omega_\rmpd|$ changes from $\gg 1$ to $\ll 1$
``suddenly'' at time $t=t_c$, at a rate much larger than $|\Omega_{\rm e}|$, 
then $\hatS$ will suddenly transition from precessing around
$\hatL_\rmd$ to precessing around $\hatL_\rmb$. The final $\thesb$ is
simply given by its value just before the transition. Depending on the
precssion phase at the transition, this final $\theta_{\rm sb,f}$ will
range from $|\theta_{\rm sd,0}-\thedb|$ to $\theta_{\rm sd,0}+\thedb$,
with $\thesd$ oscillating between $|\theta_{\rm sb,f}-\thedb|$ and
$\theta_{\rm sb,f}+\thedb$.

In the cases depicted in Section 3.3, neither the ``slow evolution''
or the ``sudden evolution'' limit applies in general, since the disc
evolution time ($\sim$~Myrs) is comparable to the disc/star precession
times for typical disc/star/binary parameters [from
eq.~(\ref{eq:Omegae}), we see that $|\Omega_e|$ reaches its minimum
value $|\Omega_\rmpd|\sin\thedb$ when $|\Omega_\rmps|= |\Omega_\rmpd
\cos\thedb|$]. Thus the evolution of $\thesd$ or $\thesb$ across the resonance
cannot be predicted in a simple analytical manner. Nevertheless 
the discussion above and Fig.~2 capture the essential features of the resonance
transition.

%This could be understood from equations (\ref{eq:Omegae}) and
%(\ref{eq:dthesd}).  We see that as the ratio
%$|\Omega_\rmps|/|\Omega_\rmpd|$ changes, $|\Omega_e|$ reaches the
%minimum value $|\Omega_\rmpd|\sin\thedb$ when $|\Omega_\rmps|=
%|\Omega_\rmpd \cos\thedb|$. Thus, the change of $\cos\thesd$
%associated with the ``resonance'' is of order $\pi \sin\theta_{\rm se}$, 
%leading to significant change in $\thesd$.

%%%%%%%%%%%%%%%%%%%%%%%%%%%%%%%%%%%%%%%%%%%%%%%%%%%%
\section{Accretion and Magnetic Torques}

In the previous sections we included only gravitational torques in the
spin-disc evolution. Here we consider the effects of mass accretion
and magnetic torques on the star. Protostars are known to have
magnetic fields of order $10^3$~G, and magnetic interactions between
the star and the disc play an important role in determining the
stellar spin evolution -- there is a large literature on this
subject; see, e.g., Bouvier et al.~(2007), Lai (2014), Romanova et
al.~(2014) for recent reviews.

%%%%%%%%%%%%%%%%%%%%%%%%%%%%%%%%%%%%
\subsection{Model for Spin and Disc Evolution}

The vast majority of studies of magnetic star - disc interactions
assume that the stellar spin axis is aligned with the disc axis.  Lai
et al.~(2011) presented a model for the evolution of 
protostellar spin for misaligned discs, including accretion, wind
and magnetic effects.  Combining equation (15) of Lai et al.~(2011)
with the gravitational torque discussed in Section 2, we have
\ba
&&{d\bS\over dt}=\lambda\,\cN_0\,\hatL_\rmd-\cN_\rms\,\hatS
+\,\cN_0\,\bn_w\cos\thesd\,\hatL_\rmd\times (\,\hatS\times\hatL_\rmd)
\nonumber\\
&&\qquad\quad +\,\cN_0\,\bn_p\cos\thesd\,\hatS\times\hatL_\rmd
+\Omega_\rmps\,\hatJ_{\rm sd}\times\bS,
\label{eq:dS}\ea
where
\be
\cN_0=\dot M\sqrt{GM_\star\rin},
\ee
is the standard accretion torque, with $\dot M$ the mass accretion
rate onto the star.

The first term on the right-hand side of equation (\ref{eq:dS})
represents the total torque in the direction of $\hatL_\rmd$; it
includes the accretion torque carried by the gas onto the star, the
magnetic braking torque associated with the disc-star linkage, as well
as any angular momentum carried away by winds/outflows from the magnetosphere
boundary. While the details are complex, all these contributions tend
to make $\lambda<1$, as suggested by numerical simulations and
semi-analytic works (see Romanova et al.~2014, Lai 2014 and references
therein). The second term in equation (\ref{eq:dS}) represents the
spindown torque along the $\hatS$ direction; for example, it could include
torques associated with winds/jets from the open field line region of
the star.
%In the aligned case ($\hatL_\rmd=\hatS$), spin equilibrium is reached when
%$\lambda \cN_0=\cN_\rms$, resulting the stellar rotation rate $\Omega_\star$
%comparable to the disc Keplerian rotation rate at $\rin$. 

The third and fourth terms in equation (\ref{eq:dS}) represent the
magnetic misalignment and precessional torques, whose physical origins
have been discussed in detail in Lai et al.~(2011) (see also Lai 1999). 
The dimensionless parameters $\bn_w$ and $\bn_p$ depend on the
details of the physics of magnetosphere-disc interaction, and are expected to be
of order unity. The last term in equation (\ref{eq:dS}) is the gravitational
torque from the disc, as calculated in Section 2.

Multiplying equation (\ref{eq:dS}) by $\hatS$ (dot product), we find that 
the magnitude of $\bS$ evolves according to
\be
{d\,S\over dt}=\cN_0\cos\thesd\left(\lambda+\bn_w\sin^2\!\thesd\right)-\cN_\rms.
\label{eq:dS2}\ee
Obviously, without a detailed knowledge of $\lambda$ (characterizing
the torques associated with accretion and magnetic star-disc linkage)
and $\cN_\rms$ (characterizing stellar outflows), it it not possible to
follow the evolution of $S$ in quantitative details.  For the reminder of
this paper, we will bypass this problem by assuming that the stellar
spin reaches the equilibrium value, such that $\Omega_\star$ is
comparable to the disc Keplerian rotation rate at $\rin$.  For our
canonical star-disc parameters [see eqs.~(\ref{eq:para1})-(\ref{eq:para2})], 
the spin ``fastness parameter'' $f_\star$ is given by equation (\ref{eq:fastness}).
Various analytical and numerical studies for aligned ($\hatL_\rmd=\hatS$)
magnetic star-disc systems suggest that 
$f_\star$ is in the range of 0.5-1 (e.g., Ghosh \& Lamb 1979; K\"onigl 1991;
Shu et al.~1994; Long et al.~2005 and references therein).

Combining equations (\ref{eq:dS}) and (\ref{eq:dS2}), we find the
stellar spin direction evolves according to the equation
\ba
&& {d\,\hatS\over dt}=\omega_0\left(\lambda-\bn_w\cos^2\!\thesd\right)
\left(\hatL_\rmd-\cos\thesd\,\hatS\right)\nonumber\\
&&\qquad\quad +\Omega_\rmps^{(m)}\,\hatJ_{\rm sd}\times\,\hatS
+\,\Omega_\rmps\,\hatJ_{\rm sd}\times\,\hatS,
\label{eq:dhatS}\ea
where
\be
\omega_0\equiv {\cN_0\over S}={\brin^{1/2}\over \bMstar}
\left(\!{0.2\over k_\star}\!\right)
\left(\!{0.1\over\bOms}\!\right)\left(\!{\dot M\over 10^{-8}M_\odot/
{\rm yr}}\!\right)\,{\rm Myr}^{-1}
\label{eq:omega0}\ee
is the inverse of the characteristic spinup time due to accretion, and 
\be
\Omega_\rmps^{(m)}=-\omega_0\,\bn_p\,\cos\thesd\left({J_{\rm sd}\over L_\rmd}\right)
\ee
is the stellar precession rate due to the magnetic torque.

The angular momentum equation for the disc can be written as
\ba
&&{d\,\bL_\rmd\over dt}=-\cN_\rmd\,\hatL_\rmd
-\,\cN_0\,\bn_w\cos\thesd\,\hatL_\rmd\times (\,\hatS\times\hatL_\rmd)
\nonumber\\
&&\qquad\quad -\,\cN_0\,\bn_p\cos\thesd\,\hatS\times\hatL_\rmd
+\Omega_\rmps\,\hatJ_{\rm sd}\times\bL_\rmd \nonumber\\
&&\qquad\quad +\Omega_\rmpd\,\hatL_\rmb\times\bL_\rmd.
\label{eq:dL}\ea
Here, the first term on the on the right-hand side represents various disc
torques (e.g., associated with angular momentum loss due to mass accretion or 
disc winds/outflows) that are aligned 
with $\hatL_\rmd$, the second and third terms are the magnetic misalignment 
and precessional torques, and the fourth and fifth terms are the gravitational
torques from the star and from the external binary (see Section 2). 
The magnitude of disc angular momentum evolves as $dL_\rmd/dt=-\cN_\rmd$, and 
the direction vector $\hatL_\rmd$ satisfies the equation
\ba
&& {d\,\hatL_\rmd\over dt}=-\left({\cN_0\over L_\rmd}\right)
\bn_w\cos\thesd \left(\hatS-\cos\thesd\,\hatL_\rmd\right)\nonumber\\
&&\qquad\quad +\Omega_\rmps^{(m)}\,\hatJ_{\rm sd}\times\,\hatL_\rmd
+\,\Omega_\rmps\,\hatJ_{\rm sd}\times\,\hatL_\rmd\nonumber\\
&&\qquad\quad +\,\Omega_\rmpd\,\hatL_\rmb\times\hatL_\rmd
\label{eq:dhatL}\ea
Combining equations (\ref{eq:dhatS}) and (\ref{eq:dhatL}), we find
\ba
&&{d\cos\thesd\over dt}=\omega_0\sin^2\!\thesd\left[\lambda-\bn_w\cos\thesd
\left(\cos\thesd+{S\over L_\rmd}\right)\right]\nonumber\\
&&\qquad\qquad +\,\Omega_\rmpd\,\hatS\cdot\left(\hatL_\rmb\times\hatL_\rmd\right).
\label{eq:dthesd2}\ea

Ignoring the term associated with disc precession [the second line of 
eq.~(\ref{eq:dthesd2})], the above equation generalizes the result
of Lai et al.~(2011), where large discs with $L_\rmd\gg S$ were
considered. The combined effects of accretion and magnetic torque on
the evolution of $\thesd$ (again ignoring disc precession) depend on the ratio
$\lambda/\bn_w$ (assuming $S/L_d\ll 1$ for simplicity; see Lai et
al.~2011, especially their Fig.~4):

(i) For $\lambda/\bn_w>1$: Regardless of the value of $\thesd$, the spin axis 
$\hatS$ is always driven towards alignment with $\hatL_\rmd$ (i.e., $\thesd$ 
always decreases).

(ii) For $\lambda/\bn_w<1$: There are two ``equilibrium'' states,
$\theta_{\rm sd+}$ and $\theta_{\rm sd-}$, given by
\be
\cos\theta_{\rm sd\pm}=\pm\sqrt{\lambda/\bn_w},
\label{eq:thesdpm}\ee
one of which ($\theta_{\rm sd+}<90^\circ$) is stable and the other 
($\theta_{\rm sd-}>90^\circ$) unstable. Thus, $\thesd$ increases toward
$\theta_{\rm sd+}$ for $\thesd<\theta_{\rm sd+}$, decreases toward
$\theta_{\rm sd+}$ for $\theta_{\rm sd+}<\thesd<\theta_{\rm sd-}$,
and increases toward $180^\circ$ for $\thesd>\theta_{\rm sd-}$.

%This shows that for $\lambda>\bn_w\cos\thesd (\cos\thesd+S/L_\rmd)$,
%mass accretion and magnetic torque tend to align $\hatS$ and $\hatL_\rmd$ 
%(for $\thesd<90^\circ$), otherwise they tend to promote misalignment
%(see Lai et al.~2011, where large discs with $L_\rmd\gg S$ were considered).

Although the values of $\lambda$ and $n_w$ are difficult to evaluate
precisely due to the complexity of magnetosphere-disc interactions, 
we expect $\lambda$ to lie in the range of 0.1-1, while $n_w$ 
ranges from somewhat less than unity to a few. Thus both cases and various
angle-dependent alignment/misalignment are possible.

In the limit of $L_\rmd\gg S$, the evolution equations for $\hatS$ and 
$\hatL_\rmd$ simplify to
\ba
&& {d\,\hatS\over dt}\simeq \omega_0\left(\lambda-\bn_w\cos^2\!\thesd\right)
\left(\hatL_\rmd-\cos\thesd\,\hatS\right)\nonumber\\
&&\qquad\quad +\left(\Omega_\rmps^{(m)}+\,\Omega_\rmps\right)
\hatL_\rmd\times\,\hatS, \label{eq:dhatS2}\\
&& {d\,\hatL_\rmd\over dt}\simeq 
\Omega_\rmpd\,\hatL_\rmb\times\hatL_\rmd\qquad\quad ({\rm for}~L_\rmd\gg S).
\label{eq:dhatL2}
\ea
We will work in this limit in the following numerical examples.

%%%%%%%%%%%%%%%%%%%%%%%%
\subsection{Numerical Results}

The various accretion/magnetic torques discussed above depend on the
parameters $\dot M$, $\lambda$, $\bn_w$ and $\bn_p$; these are in addition to
those parameters that are relevant to the gravitational torques. In
general, the accretion rate onto the star, $\dot M$, can be smaller than
the disc mass depletion rate $|\dot M_\rmd|$, since the disc may lose
mass to outflows/winds or evaporation. In the following examples, 
for simplicity, we use
\be
\dot M=|\dot M_\rmd|={M_{\rm d0}/\tau\over (1+t/\tau)^2},
\ee
and absorb the uncertainty into the parameters $\lambda$, $\bn_w$ and
$\bn_p$.

%%%%%%%%%%%%%%%%%%%%
\begin{figure}
\begin{centering}
\vskip -0.5truecm
\hskip -5truecm
\includegraphics[width=17.5cm]{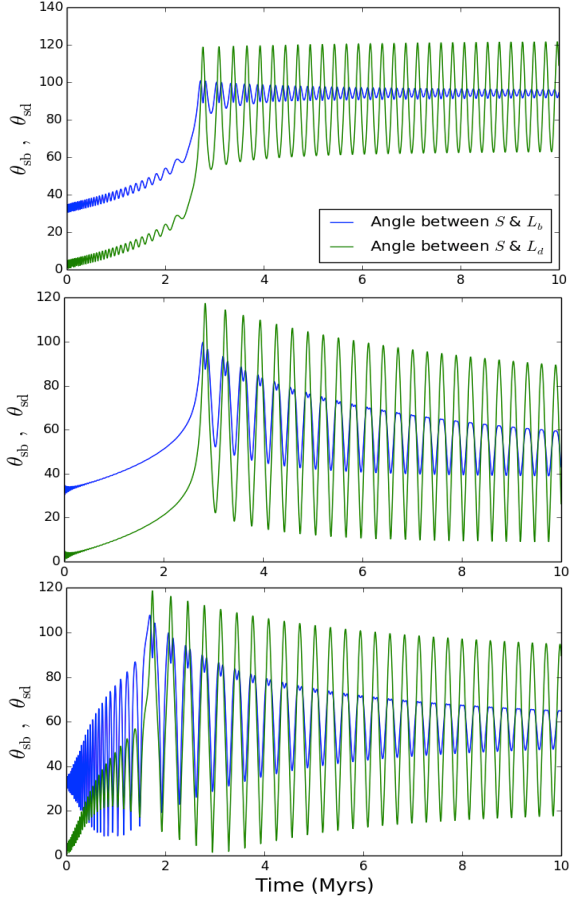}
\vskip -0.5truecm
\caption{Evolution of the stellar spin direction due to interactions
  with circumstellar disc and external binary companion, including
  accretion and magnetic torques. The
  parameters are the same as in Fig.~4 (i.e., the initial
  $\thedb=30^\circ$ and $\thesb=35^\circ$, with $a_\rmb=300$~AU,
  $\brin=1$ and $\bOms=0.1$). The upper panel is for 
  $(\lambda,\bn_w,\bn_p)=(0,0,0)$ (i.e., no accretion and magnetic
  torques), the middle panel $(1,0.5,1)$, and the bottom panel
  $(0.5,1,1)$.}
\label{fig8}
\end{centering}
\end{figure}
%%%%%%%%%%%%%%%%%%%%%

Figure \ref{fig8} shows the results of numerical integration 
of the spin axis direction for our canonical parameters 
(with $a_\rmb=300$~AU, $\thedb=30^\circ$ and $\thesb=35^\circ$),
with and without accretion/magnetic torques. For 
$(\lambda,\bn_w,\bn_p)=(1,0.5,1)$, the accretion/magnetic torques,
by themselves, tend to damp the spin-disc inclination angle $\thesd$
since $\lambda-\bn_w\cos^2\!\thesd>0$ [see eq.~(\ref{eq:dhatS2})].
In the early stage (see $t\lo 2.3$~Myrs in the middle panel of Fig.~\ref{fig8}),
these torques make the spin rotate around $\hatL_\rmb$
at the same rate as $\hatL_\rmd$ (see below for discussion).
But very quickly the gravitational effect becomes much larger than the
accretion/magnetic effects ($|\Omega_\rmpd|\gg \omega_0$ for $t\go
1$~Myrs), and $\thesd$ undergoes large variations. The fact that in the 
non-adiabatic regime, $\thesb$ is not as constant as the
zero-accretion/magnetic-torque case (compare the middle panel and the 
upper panel of Fig.~8) may seems surprising, but can be understood pictorially
from Fig.~2 and the discussion following equation (\ref{eq:dthesd2}):
As $\bS$ precesses around $\hatL_\rmb$, it is also being pulled toward 
$\hatL_\rmd$, therefore generating variations in $\thesb$ and even larger variations
in $\thesd$.

For $(\lambda,\bn_w,\bn_p)=(0.5,1,1)$, the accretion/magnetic torques
tend to increase $\thesd$ for small $\thesd$ toward $\theta_{\rm sd+}
=45^\circ$ [see eq.~(\ref{eq:thesdpm})]. This is indeed what happens
in the early stage (see the bottom panel of Fig.~\ref{fig8}). But soon
the gravitational effect takes over and $\thesd$ again exhibits large
variations. Again, the qualitative feature of this evolution can be
understood from Fig.~2 and the discussion following equation (\ref{eq:dthesd2}).

%%%%%%%%%%%%%%%%%%%%
\begin{figure}
\begin{centering}
\vskip -0.5truecm
\hskip -5truecm
\includegraphics[width=17.5cm]{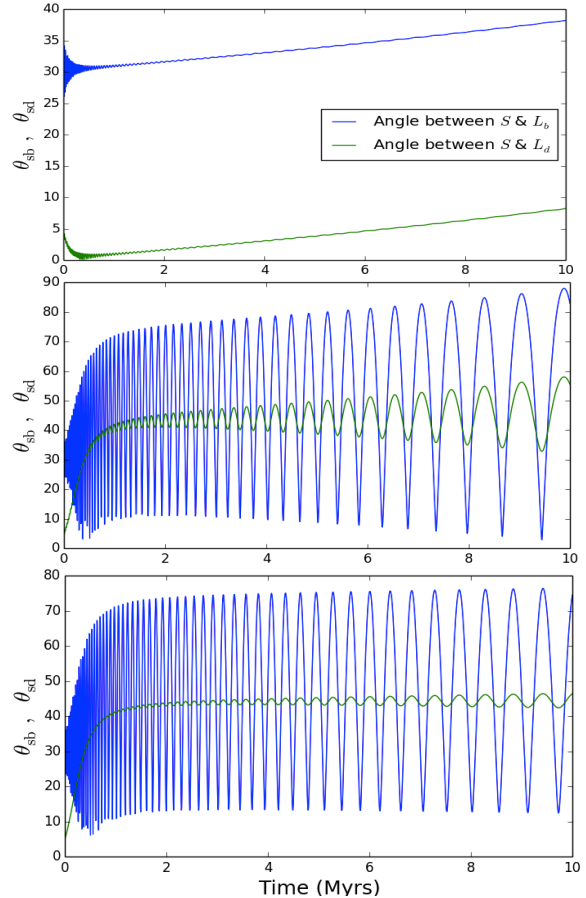}
\vskip -0.5truecm
\caption{Same as Fig.~8 except for different binary separation $a_\rmb$.
Upper panel: $a_\rmb=600$~AU, $(\lambda,\bn_w,\bn_p)=(1,0.5,1)$;
Middle panel: $a_\rmb=600$~AU, $(\lambda,\bn_w,\bn_p)=(0.5,1,1)$;
Bottom panel: $a_\rmb=1000$~AU, $(\lambda,\bn_w,\bn_p)=(0.5,1,1)$.
The other parameters are the same as in Fig.~8.}
\label{fig9}
\end{centering}
\end{figure}
%%%%%%%%%%%%%%%%%%%%%

%%%%%%%%%%%%%%%%%%%%
\begin{figure}
\begin{centering}
\vskip -0.5truecm
\hskip -6.0truecm
\includegraphics[width=16.5cm]{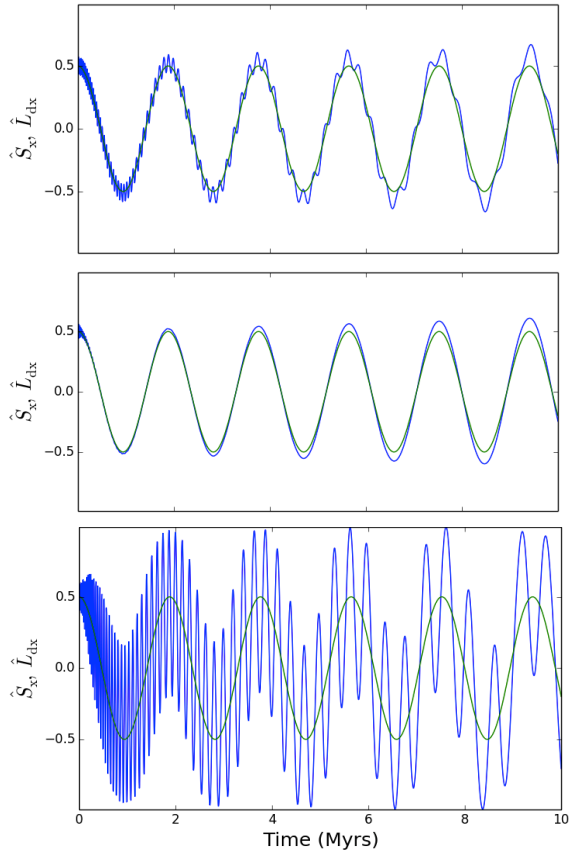}
\vskip -0.5truecm
\caption{The $x$-components of $\hatS$ (blue lines) and $\hatL_\rmd$ 
(green lines) as a function of time for models with 
$a_\rmb=600$~AU. The $z$-axis is along $\hatL_\rmb$.
The upper panel is for 
$(\lambda,\bn_w,\bn_p)=(0,0,0)$, the middle panel for
$(1,0.5,1)$ and the bottom panel for 
$(0.5,1,1)$. The other parameters are the same as in Fig.~8.
The corresponding evolutions for $\thesd$ and $\thesb$ are shown
in Fig.~7 (the lower panel) and Fig.~9 (the upper and middle panels).}
\label{fig10}
\end{centering}
\end{figure}
%%%%%%%%%%%%%%%%%%%%%

The situation is somewhat simpler for more distant binary companion
(see Fig.~\ref{fig9}). Recall that this is the case where the 
``adiabatic'' condition $|\Omega_\rmps/\Omega_\rmpd|\gg 1$ 
is always satisfied (see the lower panel of Fig.~7). 
For $(\lambda,\bn_w,\bn_p)=(1,0.5,1)$, the
accretion/magnetic torques alway act to reduce $\thesd$.  For sufficiently
large $a_\rmb$ ($\go 2000$~AU, not shown), we find that $\thesd$ indeed damps and
stays close to zero. But for smaller $a_\rmb$ (such as
$600$~AU; see the top panel of Fig.~\ref{fig9}), the damping of $\thesd$ 
is followed by its gradual growth due to the gravitational effects. This 
is simply the time-average of behaviour depicted in the lower panel of Fig.~7.
For $(\lambda,\bn_w,\bn_p)=(0.5,1,1)$ and large $a_\rmb$ (see the bottom
panel of Fig.~\ref{fig9}), the spin-disc
inclination $\thesd$ grows, driven by the magnetic torque, until it saturates
at $\theta_{\rm sd+}=45^\circ$ [eq.~(\ref{eq:thesdpm})],
where $\lambda-\bn_w\cos^2\!\thesd=0$.
For smaller $a_\rmb$ (such as 600~AU; see the middle panel), 
$\thesd$ oscillates around $45^\circ$ due to the gravitational torques.

To understand the behaviour of the $\thesd$ evolution depicted in
Fig.~\ref{fig9}, we compare in Fig.~\ref{fig10} the precession/rotation
of $\hatL_\rmd$ and $\hatS$ around $\hatL_\rmb$. Without the
accretion/magnetic torques (see the top panel), we can see that $\hatS$
precesses rapidly and follows $\hatL_\rmd$ as the latter
processes around $\hatL_\rmb$ (since $|\Omega_\rmps|\gg
|\Omega_\rmpd|$). With $(\lambda,\bn_w,\bn_p)=(1,0.5,1)$
(see the middle panel), the
damping accretion/magnetic torques suppress the precession of
$\hatS$ around $\hatL_\rmd$, making $\hatS$ ``rigidly'' follow
$\hatL_\rmd$ and precess around $\hatL_\rmb$ at rate $\Omega_\rmpd$.
This explains the smooth, close tracking of $\thesd$ and $\thesb$
(with $\thesb\simeq\thesd+30^\circ$) shown in the top panel of Fig.~\ref{fig9}
and the middle panel ($t\lo 2.3$~Myrs) of Fig.~\ref{fig8}.
On the other hand, for $(\lambda,\bn_w,\bn_p)=(0.5,1,1)$
(see the bottom panel of Fig.~\ref{fig10}), the magnetic torque
increases $\thesd$ and makes the rotation of $\hatS$ around $\hatL_\rmd$
more prominent.

%%%%%%%%%%%%%%%%%%%%
\begin{figure}
\begin{centering}
\vskip -0.5truecm
\hskip -5truecm
\includegraphics[width=17.5cm]{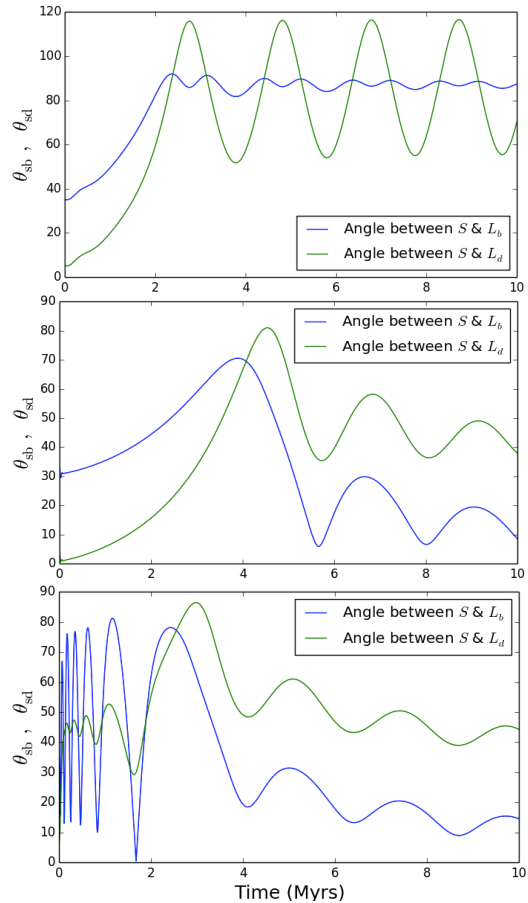}
\vskip -0.5truecm
\caption{Same as Fig.~8, for $a_\rmb=600$,
$\bar\Omega_\star=0.03$ and $\rin=2$.
Upper panel: $(\lambda,\bn_w,\bn_p)=(0,0,0)$;
Middle panel: $(1,0.5,1)$;
Bottom panel: $(0.5,1,1)$. Other parameters are the same as in Fig.~8.}
\label{fig11}
\end{centering}
\end{figure}
%%%%%%%%%%%%%%%%%%%%%

All the examples depicted above assume $\bar\Omega_\star=0.1$ and $\rin=1$.
Figure \ref{fig11} show some cases with a smaller stellar rotation rate,
$\bar\Omega_\star=0.03$, with the inner disc radius $\rin=2$, so that the 
rotation ``fastness'' parameter is approximately unchanged
[see eq.~(\ref{eq:fastness})]. This tends to increase the accretion/magnetic
effects [see eq.~(\ref{eq:omega0})], but also reduce $|\Omega_\rmps|$
[see eq.~(\ref{eq:Omegaps})]. Compare the lower panel of Fig.~7 (for 
$\bar\Omega_\star=0.1$, $\rin=1$) with the upper panel of Fig.~11, we see 
that without the accretion/magnetic torques, 
the reduced $|\Omega_\rmps|$ results is a much larger $\thesd$, with 
$\thesb$ approximately constant at later times. Including the accretion/magnetic
torques (the middle and lower panels of Fig.~11), the evolution of spin direction
is significantly modified. While some qualitative features of the evolution can be
understood from the simple ``rules'' discussed in Sections 3.1 and 3.3 (see Fig.~2),
and the discussion following equation (\ref{eq:dthesd2}), in general 
it is difficult to ``predict'' the behaviour of the spin evolution
without actual numerical integration of the evolution equations.

%%%%%%%%%%%%%%%%%%%%%%%%%%%%%%%%%%%%%%%%%%%%%%%%%%%%
\section{Viscous Damping of Disc Inclination}

In the preceding sections we have assumed that the disc precesses as a
rigid body. In reality, under the tidal forcing of the external
companion, the disc is slightly warped.  The associated viscous
dissipation tends to damp the disc axis relative to the binary
axis. Such damping has been estimated or calculated by Bate et
al.~(2000), Lubow \& Ogilvie (2000) and more recently by Foucart 
\& Lai (2014) in the case of small disc warps, based on
isotropic $\alpha$ viscosity. The result depends on $\alpha$ and
disc thickness, as well as the binary separation. However, the theory
adopted in these calculations neglects the potentially important effects of
strong, oscillating, shearing flows generated by the warp, which may
lead to the development of turbulence (Ogilvie \& Latter 2013) and 
enhance the damping rate. Thus, currently there is significant uncertainty 
about the damping timescale of disc inclination 
(see Foucart \& Lai 2014 for discussion).

Here to explore the effect of disc inclination damping on the stellar
spin evolution, we consider a simple model and assume that the
binary-disc angle decays as
\be 
\thedb (t)=\thedb(0)\exp\left(-t/t_{\rm damp}\right), 
\ee
with $t_{\rm damp}$ a free parameter. Thus, $\hatL_\rmd$ precesses
around $\hatL_\rmb$ with the frequency $\Omega_\rmpd$ but a decreasing
$\thedb$.  

\begin{figure}
\begin{centering}
\vskip -0.5truecm
\hskip -5truecm
\includegraphics[width=17.5cm]{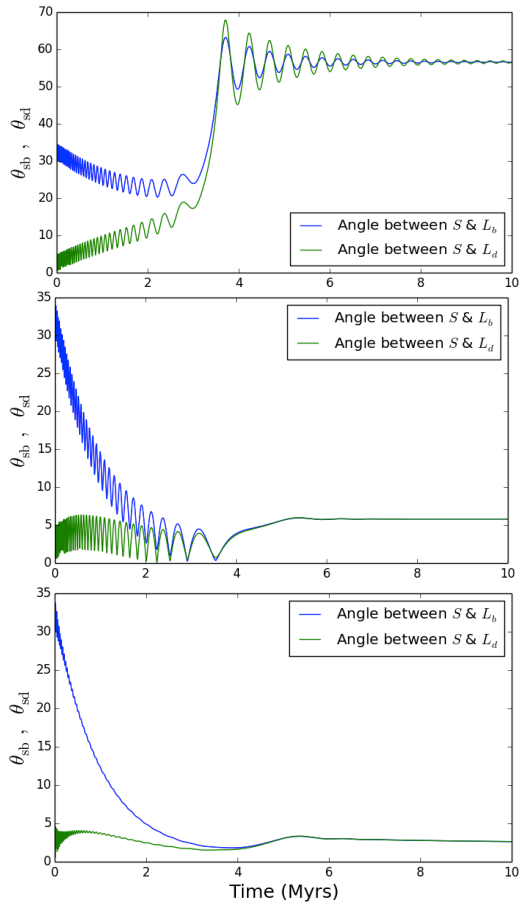}
\vskip -0.5truecm
\caption{Effect of disc inclination damping on the 
evolution of stellar spin direction. The system parameters are the
same as in Fig.~4, which corresponds to the case with no disc inclination damping
($t_{\rm damp}=\infty$), with $a_\rmb=300$~AU, $\bar\Omega_\star=0.1$,
$\rin=1$, initial $\thedb=30^\circ$, $\thesd=5^\circ$ and $\thesb=35^\circ$.
The upper panel is for $t_{\rm damp}=2$~Myrs, and the middle panel for
$t_{\rm damp}=0.8$~Myrs; both do not include accretion/magnetic torques.
The bottom panel is for $t_{\rm damp}=1$~Myrs and includes accretion/magnetic
torques, with $(\lambda,\bn_w,\bn_p)=(1,0.5,1)$.}
\label{fig12}
\end{centering}
\end{figure}

Figure \ref{fig12} shows several examples illustrating the
effect of disc inclination damping. For the canonical parameters
adopted, the damping of $\thedb$ has a significant impact on the final
spin direction if $t_{\rm damp}$ is less than a few Myrs. In
general, the final spin-disc angle $\thesd$ is reduced. For $t_{\rm damp}\lo
0.8$~Myrs, the final $\thesd$ is smaller than the initial value ($5^\circ$).
Including the spin-disc damping effect due to accretion/magnetic torques
(Section 4) further reduces $\thesd$.

%%%%%%%%%%%%%%%%%%%%
%%%%%%%%%%%%%%%%%%%%%

The examples depected in Fig.~12 (especially the middle and bottom panels)
may represent very extreme disc damping rates. They lead to an outcome at 
the end of disc evolution ($t\sim 10$~Myrs) where the star, disc and binary are
all aligned. Such an outcome is inconsistent with observations (such as
the spin-orbit misalignments in wide binaries with separations $\go 40$~AU
and the spin-disc misalignments in some young stellar objects;
see Section 1). Conceivably, these observations may be used to constrain the
physics of dissipation in warped discs.

%%%%%%%%%%%%%%%%%%%%%%%%%%%%%%%%%%%%%%%%%%%%%%%%%%%%
\section{Discussion and Conclusion}

In this paper we have examined the idea (Batygin 2012; Batygin \&
Adams 2013) that the tidal torque on a circumstellar disc from an
external binary companion can lead to the generation of misalignment
between the disc angular momentum axis and the stellar spin axis.
Such ``primordial'' spin-disc misalignment would contribute to the
spin-orbit misalignments observed in many exoplanetary systems
containing hot Jupiters. We model the star-disc-binary
interactions directly using angular momentum equations, which allow us
to describe the secular dynamics of the star-disc-binary system in a
simple way and derive the conditions (e.g., disc size, binary
separation and stellar rotation rate) under which significant
spin-disc misalignments can be produced. 

\subsection{Key Physical Effects and Results}

In general, the tidal torque from the external binary companion makes
the circumstellar disc precess around the binary angular momentum axis
(at the rate $\Omega_\rmpd$) -- the disc behaves approximately as a
rigid body because the different regions of the disc are coupled by
internal waves, viscous stresses or self-gravity.  When the disc is
misaligned with the rotation axis of the central star, it tends to
drive the stellar spin into precession around the disc axis (at the
rate $\Omega_\rmps$) due to the rotation-induced oblateness of the
star (see Fig.~1). If $|\Omega_\rmps/\Omega_\rmpd|\gg 1$, the stellar
spin axis $\hatS$ will follow the disc axis $\hatL_\rmd$
adiabatically, so that the spin-disc misalignment angle $\thesd$
remains approximately constant.  However, as the star-disc system
evolves in time (e.g., due to decreasing disc mass), $\Omega_\rmps$
decreases [see eq.~(\ref{eq:Omegaps})].  When the ratio
$|\Omega_\rmps/\Omega_\rmpd|$ crosses unity, significant spin-disc
misalignment can be generated. The importance of this secular
resonance can be understood in a geometric way (see Fig.~2 and Section
3.4). When $|\Omega_\rmps/\Omega_\rmpd|\ll 1$, the stellar spin
effectively precesses around the binary axis $\hatL_\rmb$ with
constant spin-binary inclination angle $\thesb$.

Thus, if the stellar spin and the disc axis are aligned initially,
significant spin-disc misalignment can be produced only if the system 
experiences resonance crossing during its evolution
\footnote{If the system is in the ``non-adiabatic'' regime
$|\Omega_\rmps/\Omega_\rmpd|\ll 1$ at all times, then 
then $\thesd$ will simply oscillate between $0^\circ$ and $2\thedb$,
so can periodically attain a large value if $\thedb$ is large (see Fig.~1).
However, this is unlike for typical astrophysical star-disc-binary parameters
of interest. See equation (\ref{eq:con1}).}. 
In the simple disc evolution model considered in this paper
(see Sections 2.1 and 3.3), in order to satisfy 
$|\Omega_\rmps/\Omega_\rmpd|\go 1$ at $t=0$ (with the initial disc mass
$M_{\rm di}$), we require [see eq.~(\ref{eq:ratio})]
\be
{a_\rmb\over \rout}\go 2.8\left({M_\rmb\over M_\star}\right)^{\!\!1/3}
\!\!\left({\rout\over 50\,{\rm AU}}\right)^{\!\!-1/6}
\!\left(\!{\bar\Omega_\star\over 0.1}\!\right)^{\!\!-7/9}
\!\left(\!{M_{\rm di}\over 0.1M_\star}\!\right)^{\!\!-1/3},
\label{eq:con1}\ee
where $a_\rmb$ is the binary separation, $\rout$ is the outer disc radius, 
and we have ignored the non-essential factors on the second line of 
equation~(\ref{eq:ratio}), and have assumed that $\brin$ and $\bar\Omega_\star$
are related by $f_\star=0.8$ [see eq.~(\ref{eq:fastness})] and thus
$(\bar\Omega_\star/0.1)\brin^{-2}=(\bar\Omega_\star/0.1)^{7/3}$. 
On the other hand, to satisfy $|\Omega_\rmps/\Omega_\rmpd|\lo 1$ at $t=10$~Myrs 
(with the ``final'' disc mass $M_{\rm df}$), we require 
\be
{a_\rmb\over \rout}\lo 7.6\left({M_\rmb\over M_\star}\right)^{\!\!1/3}
\!\!\left({\rout\over 50\,{\rm AU}}\right)^{\!\!-1/6}
\!\left(\!{\bar\Omega_\star\over 0.1}\!\right)^{\!\!-7/9}
\!\left(\!{M_{\rm df}\over 0.005M_\star}\!\right)^{\!\!-1/3}.
\label{eq:con2}\ee
Thus, in order for produce significant spin-disc misalignment through 
resonance crossing during the 10~Myrs of the disc lifetime (more precisely when the
disc mass changes from $0.1M_\star$ to $0.005M_\star$), the conditions
(\ref{eq:con1})-(\ref{eq:con2}) must be satisfied. Interestingly, most 
circumstellar disc systems with observed/imaged binary companions have
$a_\rmb/\rout$ in the range between a few to 10 
(e.g., Stapelfeldt et al.~1998,2003; Neuhauser et al.~2009;
Karl Stapelfeldt, private communication 2013) 
-- such a range is also largely expected from theoretical
considerations of the tidal truncation of discs in binaries 
(e.g., Artymowicz \& Lubow 1994).  So these conditions
can be met in general, although may not always. We note that even when
these conditions are not satisfied, $\thesd$ can still experience
modest growth (see, e.g., Fig.~6 and the lower panel of Fig.~7).

Another important aspect of our paper is to incorporate the effects of
accretion and magnetic torques in the evolution of the stellar spin
direction (Section 4). In general, the magnetic torques between a
protostar and its circumstellar disc can induce spin-disc misalignment
and contribute to the star-disc mutual precession (Lai et al.~2011),
while accretion tends to damp the misalignment.  Unfortunately, given
the intrinsic complexity of the magnetosphere-disc interaction
physics, the net effect accretion and magnetic torques cannot be
determined with certainty at present. Nevertheless, we show that these
torques in general can influence the evolution of the stellar spin axis 
in a significant way (see Figs.~8,9,11).

Finally we briefly explore the effect of disc inclination damping
(relative to the binary) associated with viscous dissipation of disc
warps (Section 5).  Only with extreme damping timescales (less than 1
Myrs) can the spin-disc misalignments be significantly affected.  By
comparing with the observations of spin-orbit misalignments in wide
binaries and spin-disc misalignments in protostellar systems, it may be
possible to constrain the physics of disc warp dissipations.

\subsection{Implications}

The general implication of our work is that in the presence of a
binary companion, stellar spin-disc misalignments can be generated
within the typical lifetimes of protoplanetary discs under a wide
range of conditions, as long as the binary axis is somewhat misaligned
with the disc axis (by more than a few degrees). There is ample
evidence of such disc-binary misalignments from observations of
protostellar jets or direct imagings of protostellar discs (see
Section 1). In fact, for very reasonable binary/disc parameters, it
has proven difficult to avoid the production of significant spin-disc
misalignments from small initial values.  Although challenging, it
would be of great interest to measure or constrain the spin-disc
inclination angles of young protostellar disc systems (rather than
debris disc systems; see below).

Our general conclusion is consistent with the observations of
spin-orbit misalignments in wide ($a_\rmb \go 40$~AU) binaries (Hale
1994). Since there is significant evolution of the spin-disc
misalignment angle during the lifetime of the disc, our work suggests
that the observed spin-orbit misalignments do not necessarily
translate into a large difference in the angular momentum directions of
separate molecular cloud cores that form the binary components.

Concerning giant planets, our work suggests that spin-disc
misalignments generated by star-disc-binary interactions 
can make a significant contribution to the observed 
spin-orbit misalignments in hot Jupiter systems. 
Such misalignments may already be present prior to any subsequent
(after disc dispersal) dynamical processes. Indeed, the same kind of 
inclined binaries are invoked in the Lidov-Kozai mechanism for producing
hot Jupiters (see references in Section 1). Moreover, in order 
for the Lidov-Kozai migration to operate on a planet at the semi-major axis $a_p$,
the Kozai oscillation period ($\simeq 2\pi/\dot\omega_{\rm Kozai}$)
\be
P_{\rm Kozai}\simeq 10^6\left({a_\rmb\over 500\,{\rm AU}}\right)^3
\left({a_p\over 5\,{\rm AU}}\right)^{\!\!-3}\,{\rm yrs}
\ee
must be shorter than the apsidal precession period of the planetary
orbit due to general relativity (e.g., Holman et al.~1997). That is, the ratio
\be
{\dot\omega_{\rm GR}\over \dot\omega_{\rm Kozai}}\simeq 6\times 10^{-3}
\left(\!{M_\star\over M_\rmb}\!\right)\left({a_\rmb\over 500\,{\rm AU}}\right)^{\!3}
\left({a_p\over 5\,{\rm AU}}\right)^{\!\!-4}
\ee
must be less than unity. Thus, if the planet was formed at $a_p\lo
1.4$~AU or had migrated (due to disc-driven migration) to such a
distance by the end of the protoplanetary disc phase, the binary
companion (assuming $a_\rmb =500$~AU) would not be able to induce any
Kozai oscillation. Nevertheless, such ``failed-Kozai'' systems may
still have significant spin-orbit misalignments because of the
primordial spin-disc misalignments studied in this paper.  Of course,
whether such systems can have any significant eccentricities or can
become hot Jupiters is a different question.
In this regard, it is of interest to note that there appears to be 
a significant lack of super-eccentric proto-hot Jupiters in the Kepler
sample (Dawson et al.~2012), suggesting that many hot Jupiters are not 
produced by high-eccentricity migration mechanism 
(see Socrates et al.~2012; Dawson \& Murray-Clay 2013). 

To predict the planet's orbital inclination at the end of the
protoplanetary disc phase, it will be necessary to keep track of the
evolution of the planet's angular momentum (both magnitude and
direction) as the disc evolves and precesses, together with all the
other effects considered in this paper. Planet-disc interaction will
be an important part of the story (e.g., Baruteau et al.~2013; see
also Marzari \& Nelson 2009; Terquem 2013; Xiang-Gruess \& Papaloizou
2013).  Presumably, when the disc is sufficiantly massive as the
planet forms and evolves, the angular momentum of the planet will be
strongly coupled to that of the disc. If the planet migrates to the
close vicinity of the star, the mutual torque between the star and the
planet must also be considered.  We plan to address some of these
issues in the future.

Concerning debris disc systems, current observations do not reveal any
evidence of significant spin-disc misalignments (Watson et al.~2011;
Greaves et al.~2013; Kennedy et al.~2013). This suggests that the
observed/measured systems, so far, do not have binary companions at
all or companions with appropriate separations. Finding binary
companions in debris disc systems and measure/constrain the relative
orientations of spin, disc and binary axes would be of great
interest. On the theoretical side, it would be necessary to extend
our study to much later times (beyond 10~Myrs).  This would require
more reliable models for the evolution of discs from the
protoplanetary phase to the debris disc phase.

\section*{Acknowledgments}
I thank Diego Munoz for teaching me PYTHON, Kassandra Anderson,
Francois Foucart, Diego Munoz and Karl Stapelfeldt for discussions,
and Yanqin Wu for comments. This work has been supported in
part by NSF grants AST-1008245, AST-1211061 and NASA grant NNX12AF85G.

%\bibliographystyle{mn.bst}
%\bibliography{pmodes}

%%%%%%%%%%%%%%%%%%%%%%

\end{document}